\documentclass[aip,reprint,superscriptaddress,longbibliography,floatfix]{revtex4-1}

\usepackage{amsmath,braket}
\usepackage{graphicx}
\usepackage{mathtools}
\usepackage{color}
\usepackage{xcolor}
\usepackage{setspace}
\usepackage{float}
\usepackage{hyperref}

\makeatletter
\newcommand{\printfnsymbol}[1]{%
  \textsuperscript{\@fnsymbol{#1}}%
}
\makeatother

\begin{document}
\title{Suppression of surface-related loss in a gated semiconductor microcavity}

\author{Daniel Najer}
\thanks{Daniel Najer and Natasha Tomm contributed equally to this work}
\affiliation{Department of Physics, University of Basel, Klingelbergstrasse 82, CH-4056 Basel, Switzerland}

\author{Natasha Tomm}
\affiliation{Department of Physics, University of Basel, Klingelbergstrasse 82, CH-4056 Basel, Switzerland}

\author{Alisa Javadi}
\affiliation{Department of Physics, University of Basel, Klingelbergstrasse 82, CH-4056 Basel, Switzerland}

\author{Alexander R.\ Korsch}
\affiliation{Lehrstuhl f\"{u}r Angewandte Festk\"{o}rperphysik, Ruhr-Universit\"{a}t Bochum, D-44780 Bochum, Germany}

\author{Benjamin Petrak}
\affiliation{Department of Physics, University of Basel, Klingelbergstrasse 82, CH-4056 Basel, Switzerland}

\author{Daniel Riedel}
\affiliation{Department of Physics, University of Basel, Klingelbergstrasse 82, CH-4056 Basel, Switzerland}

\author{Vincent Dolique}
\affiliation{Laboratoire des Mat\'{e}riaux Avanc\'{e}s (LMA), IN2P3/CNRS, Universit\'{e} de Lyon, F-69622 Villeurbanne, Lyon, France}

\author{Sascha R.\ Valentin}
\affiliation{Lehrstuhl f\"{u}r Angewandte Festk\"{o}rperphysik, Ruhr-Universit\"{a}t Bochum, D-44780 Bochum, Germany}

\author{R\"{u}diger Schott}
\affiliation{Lehrstuhl f\"{u}r Angewandte Festk\"{o}rperphysik, Ruhr-Universit\"{a}t Bochum, D-44780 Bochum, Germany}

\author{Andreas D.\ Wieck}
\affiliation{Lehrstuhl f\"{u}r Angewandte Festk\"{o}rperphysik, Ruhr-Universit\"{a}t Bochum, D-44780 Bochum, Germany}

\author{Arne Ludwig}
\affiliation{Lehrstuhl f\"{u}r Angewandte Festk\"{o}rperphysik, Ruhr-Universit\"{a}t Bochum, D-44780 Bochum, Germany}

\author{Richard J.\ Warburton}
\email{richard.warburton@unibas.ch}
\affiliation{Department of Physics, University of Basel, Klingelbergstrasse 82, CH-4056 Basel, Switzerland}

\date{\today}

\begin{abstract}
We present a surface passivation method that reduces surface-related losses by almost two orders of magnitude in a highly miniaturized GaAs open microcavity. The microcavity consists of a curved dielectric distributed Bragg reflector (DBR) with radius $\sim 10$\,$\mu$m paired with a GaAs-based heterostructure. The heterostructure consists of a semiconductor DBR followed by an n-i-p diode with a layer of quantum dots in the intrinsic region. Free-carrier absorption in the highly doped n- and p-layers is minimized by positioning them close to a node of the vacuum electromagnetic-field. The surface, however, resides at an anti-node of the vacuum field and results in significant loss. These losses are much reduced by surface passivation. The strong dependence on wavelength implies that the main effect of the surface passivation is to eliminate the surface electric field, thereby quenching below-bandgap absorption via a Franz-Keldysh-like effect. An additional benefit is that the surface passivation reduces scattering at the GaAs surface. These results are important in other nano-photonic devices which rely on a GaAs-vacuum interface to confine the electromagnetic field.
\end{abstract}

\maketitle
\section{Introduction}
Concepts in cavity quantum-electrodynamics (QED) can be implemented using semiconductors. A semiconductor based microcavity can be created using a micropillar\cite{Reithmaier2004,Somaschi2016,Ding2016}, a photonic crystal cavity \cite{Yoshie2004,Kuruma2020}, a whispering-gallery resonator~\cite{Guha2017}, and an open microcavity~\cite{Barbour2011,Greuter2015,Najer2019}. Quantum dots within these structures mimic atoms. In the limit of a single quantum dot (QD), a single-photon source can be realized by exploiting the weak-coupling regime of cavity-QED\cite{Senellart2017}. The strong-coupling regime of cavity-QED has been accessed with three different microcavity platforms~\cite{Reithmaier2004,Yoshie2004,Najer2019}.

In all these semiconductor-based applications of cavity-QED, minimizing the absorption and scattering losses in the microcavity is very important. For single-photon sources operating in the weak-coupling regime, an efficient photon extraction from the microcavity is essential~\cite{Somaschi2016,Ding2016}. In the strong-coupling regime, a coherent exchange between an exciton in the QD and a photon in the micro-cavity is only possible if the exciton-photon coupling exceeds the rate of photon loss. Typically, this requires the development of low-mode volume, high $\mathcal{Q}$-factor microcavities. A recurring theme in the development of such microcavities is the role of the GaAs surface. At the semiconductor surface, the symmetry of the lattice is broken. The GaAs surface is quite complex -- there are a number of possible surface reconstructions, and a thin oxide layer typically forms on exposure to air~\cite{Demanet1985}. Reducing surface-related absorption is crucial in the development of GaAs-based cavity-QED.

Recently, considerable success in implementing cavity-QED was reported with a QD in an open microcavity~\cite{Barbour2011,Greuter2015,Najer2019,Tomm2020}. The ``bottom" mirror is a semiconductor distributed Bragg reflector (DBR), the ``top" mirror a curved, dielectric DBR. The position of the bottom mirror can be tuned {\em in situ} with respect to the top mirror, allowing a single QD to be brought into resonance with the microcavity mode. In the latest developments, the QDs are embedded in an n-i-p diode~\cite{Najer2019,Tomm2020}. Both n- and p-doped GaAs result in free-carrier absorption~\cite{Casey1975}. To minimize the absorption within the microcavity, the n- and p-doped layers are made as thin as possible and are positioned close to the node of the vacuum electric field. This technique, positioning an absorbing layer at a vacuum field node, can also be used to reduce losses at the GaAs surface. But in this case, it involves a serious compromise. If there is a node at the surface, the largest vacuum electric field lies in the vacuum-gap and not in the GaAs material on account of interferences in the device. This reduces considerably the coupling of a QD to the vacuum electric field. An acceptably large coupling is only possible if there is a vacuum field anti-node at the surface. Success with the n-i-p devices in an open microcavity was only possible after passivating the surface~\cite{Najer2019,Tomm2020}. The role played by the passivation is elucidated here.

\begin{figure}[t!]
\centering
\includegraphics[width=\columnwidth]{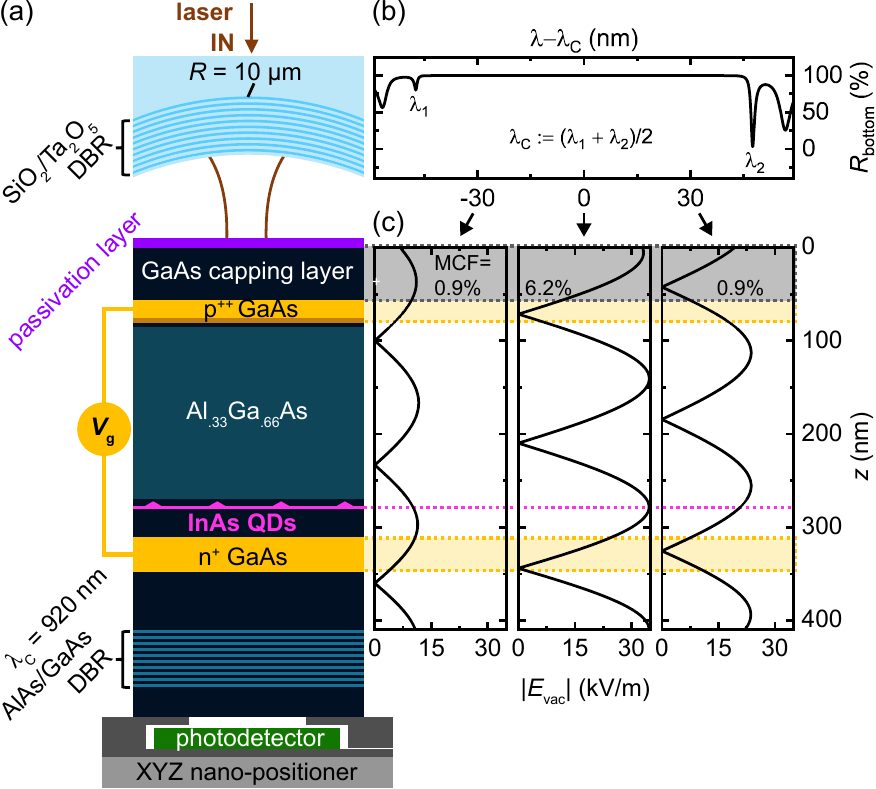}
\caption{Ultrahigh $\mathcal{Q}$-factor optical microcavity as sensitive probe of surface-related absorption. (a) Schematic of the microcavity involving a curved dielectric-DBR and an n-i-p heterostructure with self-assembled InAs QDs on top of a semiconductor DBR (``nip-DBR''). (b) Simulated reflectance of the nip-DBR with stopband (SB) center $\lambda_{\rm C}=920$\,nm. (c) Calculated vacuum-field amplitude across the heterostructure for three different wavelengths ($-30, 0, +30$)\,nm with respect to the SB center. As the antinodes of the vacuum-field shift in position with wavelength thereby changing the modal confinement factor (MCF) in the GaAs capping layer, surface-related absorption in the capping layer ($10^{-10}$--$10^{-8}$\,cm$^{-1}$) can be probed via the microcavity by measuring its $\mathcal{Q}$-factor across the SB. Note that at $\lambda_{\rm C}$, where the coupling to the QDs is maximized, free-carrier absorption in the highly doped p- and n-gates is minimized by placing them close to a vacuum-field node. Note also that the highly reduced vacuum-field at $\lambda-\lambda_{\rm C}=-30$\,nm arises as at this wavelength, the largest vacuum electric-field is located in the vacuum gap.}
\label{design2}
\end{figure}

We probe the surface-related absorption in a GaAs open microcavity. The main diagnostic tool is a measurement of the wavelength dependence of the $\mathcal{Q}$-factor. With an untreated surface, we find that the $\mathcal{Q}$-factor is modest, approximately $10^{4}$ at the stopband center, much lower than the value expected from the mirror designs. Following surface passivation, we find that the $\mathcal{Q}$-factor increases to $\simeq 10^{6}$ at the stopband center, close to the value expected from the mirror designs. This shows that, first, the dominant loss mechanism in the untreated case is related to the GaAs surface, and second, that surface passivation remedies this loss. For the untreated surface, the $\mathcal{Q}$-factor has a very strong dependence on wavelength, increasing rapidly on tuning to lower wavelengths. By comparing the $\mathcal{Q}$-factor to the result of model calculations, we find that we can account quantitatively for the $\mathcal{Q}$-factor by ascribing the loss to absorption in the capping layer, the final 55-nm-thick GaAs layer of the heterostructure. The absorption in the capping layer is an exponential function of the photon energy, pointing to Franz-Keldysh-like absorption induced by a strong electric field at the surface~\cite{Franz1958,Keldysh1957,Aspnes1966,Hader1997,Knupfer1993}. In turn, this demonstrates the main role of the surface passivation layer in this device: it reduces the surface electric field, thereby much reducing the Franz-Keldysh (F-K) absorption. The standard analytic result for the F-K effect describes the absorption at the unpassivated surface but with an electric field much higher than in the standard picture (mid-gap pinning).

\section{The open, tunable microcavity}

The microcavity~\cite{Barbour2011,Greuter2014} consists of a curved dielectric DBR -- the template is produced by CO$_2$-laser ablation~\cite{Hunger2012} -- paired with an ``nip-DBR" semiconductor heterostructure. The InAs QDs are embedded in the intrinsic part of the n-i-p diode; the diode resides on top of a semiconductor DBR, Fig.~\ref{design2}(a). We employ two dielectric top DBRs. The first (DBR-I) is composed of 22 pairs of SiO$_2$($\lambda/4$) and Ta$_2$O$_5$($\lambda/4$) (where $\lambda$ depicts the wavelength in each material) and is terminated with SiO$_2$. The stopband (SB) center\footnote{Note that in this work, we define the SB center as the mean value of the two wavelengths at the local minima (with $R<90\%$) of the calculated reflectance spectrum that are closest to the maximum mirror reflectance (Fig.~\ref{design2}(b)).} is 973\,nm. The second (DBR-II) is composed of 15 pairs of SiO$_2$($\lambda/4$) and Ta$_2$O$_5$($\lambda/4$), Ta$_2$O$_5$ terminated, and has its SB center at 930\,nm. The semiconductor DBR consists of 46 pairs of AlAs($\lambda/4$) and GaAs($\lambda/4$). The heterostructure is a $1.5\lambda$-layer of GaAs including doped layers acting as top-gate (p$^{++}$, $10^{19}$ cm$^{-3}$) and back-gate (n$^{+}$, $2 \cdot 10^{18}$ cm$^{-3}$). The QD layer is placed at an antinode of the vacuum electric field (at a distance $\lambda$ below the surface). The intrinsic region between QDs and back-gate acts as tunnel barrier for electrons and ensures that the QDs operate under Coulomb blockade at low temperature~\cite{Warburton2000}. Using a piezo-based XYZ nano-positioner, the microcavity features full {\em in situ} tunability at cryogenic temperatures. 

A measurement of the $\mathcal{Q}$-factor across the SB of the nip-DBR (Fig.~\ref{design2}(b)) reveals possible sources of loss in the heterostructure due to the fact that the standing wave inside the cavity shifts with wavelength (Fig.~\ref{design2}(c)). For instance, losses in the capping layer depend on the exact wavelength: at a wavelength-detuning of ($-30,0,30$)\,nm with respect to the nip-DBR's SB center, the calculated modal confinement factor (MCF -- defined as the vacuum electromagnetic-energy confined in the layer-of-interest divided by the zero-point energy, $\hbar\omega/2$, i.e.\ the total energy of the vacuum-field mode.) of the capping layer is ($0.9\%,6.2\%,0.9\%$), respectively. Therefore, if the dominant loss mechanism in the microcavity takes place within the capping layer then the change in MCF will result in a strong dependence of the $\mathcal{Q}$-factor across the stopband. Furthermore, by characterizing the mirrors carefully and by simulating the entire structure with transfer-matrix calculations, measurements of the $\mathcal{Q}$-factor not only reveal the location of the dominant loss process but can also be used to determine the loss quantitatively. 

\section{G\lowercase{a}A\lowercase{s} surface passivation}
Surface passivation of GaAs replaces the native oxide with a thin Al$_2$O$_3$ layer\cite{Guha2017,Liu2018}. The surface passivation recipe follows in part one of the procedures described in Ref.~\cite{Xuan2007}. As a first cleaning step, the processed semiconductor sample (already containing Au contact pads) is successively immersed in acetone, isopropanol and ethanol inside an ultrasonic bath at $T=40\,^\circ$C. To prevent surface treatment of the contact pads, they are covered by a manually applied drop of photoresist (AZ1512HS, Microchemicals GmbH) and baked for 10 min\ at $T=100\,^\circ$C. At room temperature, the sample is dipped into an HCl solution (25\%) for 1 min in order to remove the native oxide~\cite{Xuan2007,Rebaud2015}. The sample is then rinsed with deionized water for $\sim1$\,s and immediately immersed in an (NH$_4$)$_2$S solution (20\%) for 10 min. This procedure passivates the surface with sulfur~\cite{Yablonovitch1987,Ohno1991}, preventing the native oxide from reforming. The S-layer is however not robust. For this reason, it is removed and replaced with an Al$_2$O$_3$ layer. To achieve this, on emerging from the (NH$_4$)$_2$S solution, the sample is blown dry with nitrogen and immediately transferred into an atomic-layer deposition (ALD) chamber (Savannah 100, Cambridge NanoTech Inc.). 

\begin{figure}[t!]
\centering
\includegraphics[width=\columnwidth]{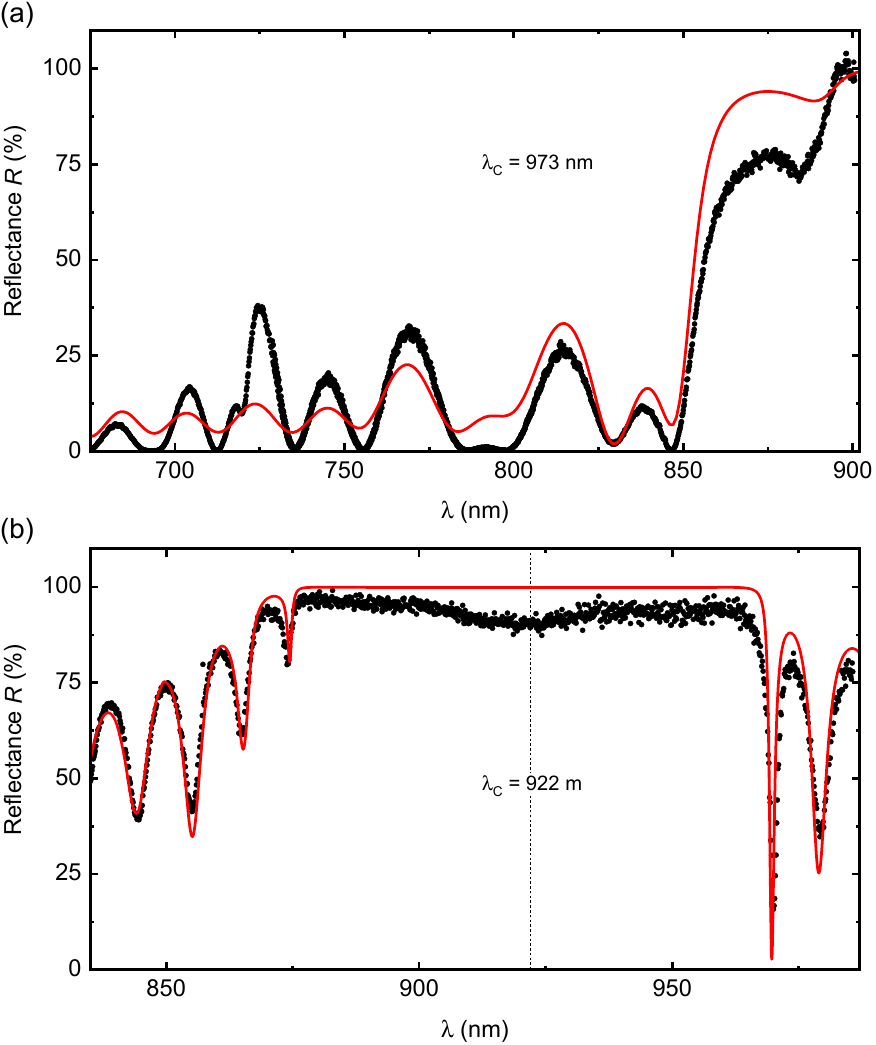}
\caption{Mirror characterization via reflection measurements. Each mirror is investigated by recording the spectrum of white light reflected off the sample using a dark-field confocal microscope~\cite{Kuhlmann2013}. Via 1D transfer matrix methods (Essential Macleod), the designed layer thicknesses can be refined to fit the experimentally observed oscillations. (a) Dielectric DBR-I. The reflected signal is recorded on a flat surface away from the curved part of the mirror and normalized by the white-light spectrum. (b) Unpassivated nip-DBR. Here, the reflectance spectrum is obtained by normalizing the reflected signal from the mirror by the reflected signal from a Au contact pad (by moving the piezo-nanopositioner laterally by a few microns).}
\label{mirrorcharac}
\end{figure}

The following ALD recipe is chosen to deposit $\sim 8$\,nm of Al$_2$O$_3$ onto the sample surface: $T=150\,^\circ$C, first pulse 50 ms (water), wait 12 s, second pulse 40 ms (TMA), wait 10 s; the cycle is repeated 80 times. The Al$_2$O$_3$ layer acts as diffusion barrier for oxygen~\cite{Robertson2015}, thus preventing reoxidation of the etched GaAs surface.

After surface passivation, the remaining challenge is to remove the photoresist that has been cross-linked due to the high temperature, $T=150\,^\circ$C, inside the ALD chamber. The use of \emph{N}-methyl-2-pyrrolidone (NMP) at elevated temperatures was shown to remove successfully  the cross-linked photoresist. The sample is immersed in NMP for 9--20$\,$h (20$\,$h yielded a better result) at $T=40\,^\circ$C and then successively cleaned for 5 min in NMP, acetone, isopropanol and methanol inside an ultrasonic bath at $T\sim 56\,^\circ$C. As a final step, a polymeric strip coating (First Contact, Photonic Cleaning Technologies) is used to remove any final residues from the sample surface.

\section{Individual mirror characterization}
\label{mirrocharac}

Each DBR has a high reflectance for wavelengths within the stopband (SB). Outside the stopband, there are oscillations in the reflectivity as a function of wavelength. These oscillations are sensitive to the exact layer thicknesses in the particular mirror -- this dependence is exploited to characterize the layers in each DBR. 

As depicted in Fig.~\ref{mirrorcharac}, the cavity's top and bottom mirrors are characterized at $T=4.2$\,K by a broadband light-source (white light-emitting-diode or halogen lamp) and a dark-field confocal microscope~\cite{Kuhlmann2013}. The light from the source is coupled into a single-mode optical fiber, the output of which is collimated and focussed onto the sample surface with an objective lens (NA=0.55). Cross-polarizing elements are used in the beam path to reject all but the light reflected from the sample surface. The detection fiber is connected to a spectrometer~\citep{Kuhlmann2013}. The reflected light from a metallic mirror, the Au contact pad in the case of the nip-DBR, is used to record a reference spectrum. The nip-DBR's reflectance spectrum is obtained by dividing its reflected spectrum by the reference spectrum. Due to the absence of a metallic reference surface on top of the dielectric DBR-I in Fig.~\ref{mirrorcharac}(a), an exponential fit of the reference spectrum from Fig.~\ref{mirrorcharac}(b) is used instead and the maximum reflectance is normalized to one.

Via 1D transfer matrix methods (Essential Macleod, Thin Film Center Inc.) the design layer thicknesses can be refined in order to fit the reflectivity oscillations outside the stopband. The obtained models for each DBR (red solid lines in Fig.~\ref{mirrorcharac}) provide a convincing description of these oscillations. These mirror descriptions are then used to simulate the cavity performance, in particular the $\mathcal{Q}$-factors and transmittance values at resonance as a function of the vacuum-gap between the mirrors (Sec.~\ref{macleod}). The slight discrepancy between experiment and model arises from the difficulty of recording precisely a reference spectrum for the white-light source.

\section{Microcavity characterization: $\mathcal{Q}$-factors}

\begin{figure}[H]
\centering
\includegraphics[width=\columnwidth]{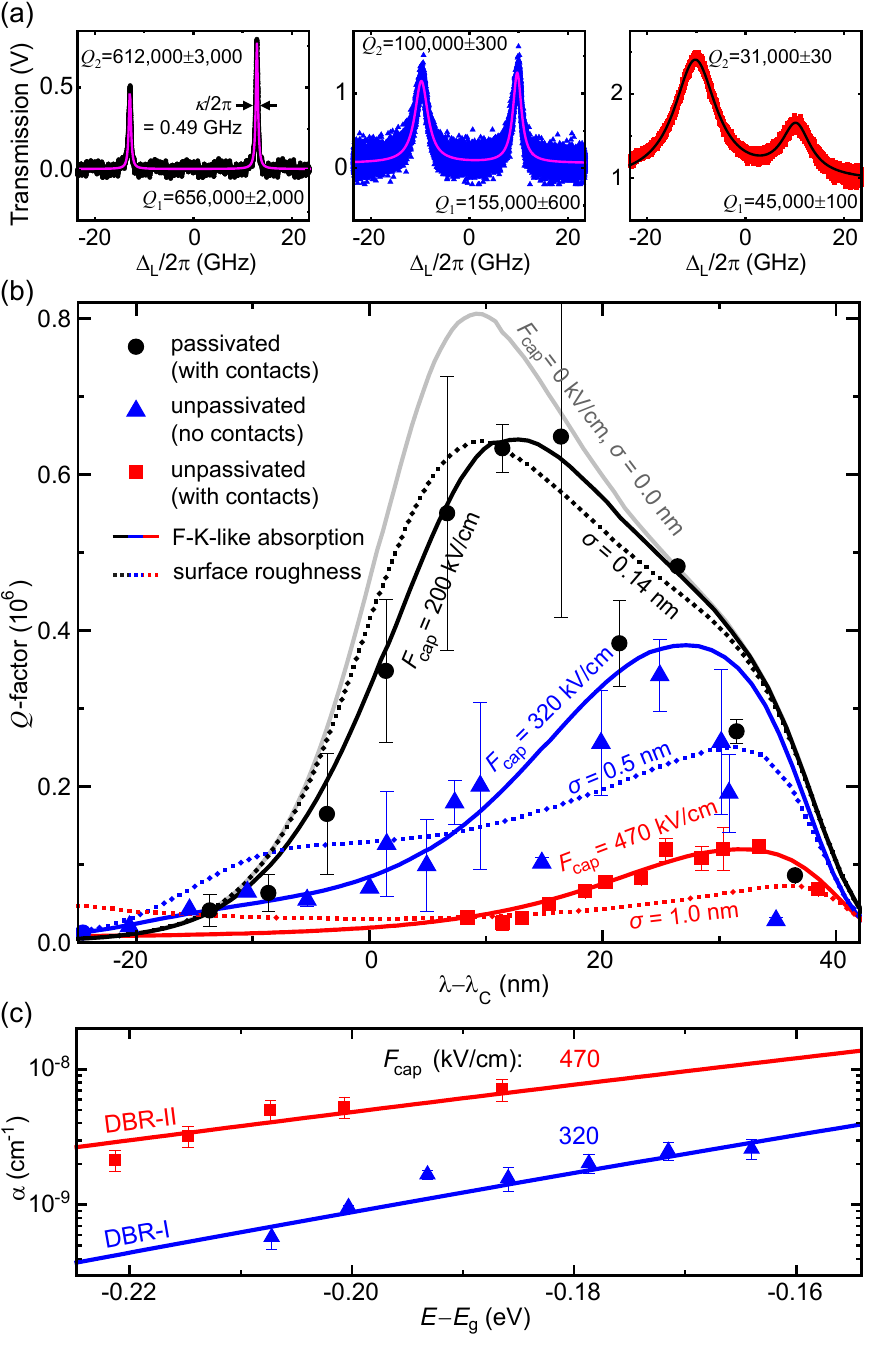}
\caption{Microcavity characterization via $\mathcal{Q}$-factor measurements using: DBR-I paired with passivated nip-DBR with contacts (black dots), two unpassivated nip-DBRs without contacts (blue triangles); DBR-I (II) paired with five (one) unpassivated DBRs with contacts (red squares). (a) Measured transmission signal as a function of laser detuning at $\lambda-\lambda_C\sim10$\,nm at fixed mirror separation. The $\mathcal{Q}$-factor is determined by a double-Lorentzian fit (solid lines). (b) Evaluated $\mathcal{Q}$-factors for several wavelengths. Mean values and standard deviations originate from data from two cavity modes, and up to six measurements. Two surface-loss mechanisms are modeled: absorption and scattering. The solid lines are calculated $\mathcal{Q}$-factors taking into account free-carrier absorption (absorption coefficients from Ref.\cite{Casey1975}) and F-K absorption\cite{Hader1997,Knupfer1993} for electric fields in the capping layer $F_{\rm cap}=(200,320,470)$\,kV/cm. The dotted lines are calculated taking into account surface scattering ~\cite{Carniglia2002} only. For the passivated case, the black dotted line is modelled with a roughness at both the GaAs-alumina and alumina-vacuum interfaces $\sigma = (\sigma_{\rm GaAs-Al2O3},\sigma_{\rm Al2O3-vac})$, and found to be $\sigma=(0.14,0.14)$\,nm; alternatively $\sigma=(0.0,0.55)$\,nm yields similar results. In the unpassivated cases, the blue and red dotted lines are modelled with $\sigma_{\rm GaAs-vac}=0.5$\,nm and $\sigma_{\rm GaAs-vac}=1.0$\,nm at the GaAs-vacuum interface, respectively. Gray solid line is model without any surface losses. (c) By comparing measured and simulated $\mathcal{Q}$-factors, and assuming that the scattering losses are negligible, the absorption coefficient $\alpha$ can be deduced as a function of photoenergy. $\alpha$ is fitted to the F-K result using $F_{\rm cap}$ as fitting parameter.}
\label{cavitycharac}
\end{figure}

A microcavity is constructed using passivated and unpassivated semiconductor-DBRs and a curved dielectric DBR (radius of curvature $\sim 7$--$16$\,$\mu$m) similar to the ones characterized in Fig.~\ref{mirrorcharac}. Via narrowband-laser transmission-measurements, each microcavity is characterized by determining its $\mathcal{Q}$-factor across the SB of the semiconductor DBR. The transmission signal is measured as a function of laser frequency keeping the cavity length fixed (Fig.~\ref{cavitycharac}(a)). To change the cavity's resonance frequency, the mirror separation is changed by means of the Z nano-positioner. 

A $\mathcal{Q}$-factor is obtained for every pair of longitudinal (TEM$_{00}$) modes at the minimum mirror separation ($\sim 2$--$4$\,$\mu$m, depending on wavelength and mirror-crater depth~\cite{Greuter2014}) by fitting a double-Lorentzian. Fig.~\ref{cavitycharac}(a),(b) show the results for an electrically-contacted passivated sample\footnote{Note that the measured $\mathcal{Q}$-factors obtained with an electrically-contacted passivated sample and with an electrically-uncontacted passivated sample were similar. The latter are not shown in Fig.~\ref{cavitycharac}(b).} (black dots), an electrically-uncontacted unpassivated sample (blue triangles) as well as an electrically-contacted unpassivated sample (red squares), paired with DBR-I as top mirror. Without passivation, the $\mathcal{Q}$-factor is around $10^{5}$ for the electrically-uncontacted unpassivated sample at the SB center, and too low to measure precisely for the electrically-contacted unpassivated sample. At a red-shift of 10\,nm from the SB center, the $\mathcal{Q}$-factor for the electrically-contacted unpassivated sample is on average $3.8\cdot 10^{4}$, and $1.28\cdot 10^{5}$ for the electrically-uncontacted unpassivated sample (Fig.~\ref{cavitycharac}(a).). These values are much smaller than the values expected from the DBRs -- they signify that there is a significant loss mechanism. Around the SB center, where the coupling to the QD layer is maximized ($\lambda_{\rm C}=915$--925\,nm), the $\mathcal{Q}$-factors are strongly decreased by this loss mechanism -- the loss impacts significantly the performance of a QD in the microcavity.

As the wavelength approaches the red-end of the SB, the $\mathcal{Q}$-factors for the unpassivated samples increase. There is a pronounced asymmetry with respect to the SB center: the $\mathcal{Q}$-factors fall monotonically as the wavelength shifts to the blue with respect to the SB center. These results imply that the loss mechanism is a strong function of wavelength. 

After passivation, the $\mathcal{Q}$-factor around the SB center increases. At a red-shift of 10\,nm from the SB center, the $\mathcal{Q}$-factor increases to a very large value, $6.34\cdot 10^{5}$ (Fig.~\ref{cavitycharac}(a)). Furthermore, the dependence of the $\mathcal{Q}$-factor on wavelength is much more symmetric with respect to the SB center. In fact, the decrease on the blue-side reflects the decrease in reflectivity of the dielectric mirror which, for fabrication reasons, has a maximum reflectivity at a wavelength of 973\,nm.

The passivation procedure changes the properties of the surface but leaves the rest of the microcavity unchanged. The drastic increase of the $\mathcal{Q}$-factors after surface passivation leads therefore to the conclusion that the losses limiting the $\mathcal{Q}$-factors of unpassivated microcavities are related to the semiconductor surface. Specifically, the loss, either an absorption or scattering mechanism, originates at the GaAs surface itself or in the GaAs layer immediately below the surface.

\begin{figure}[t!]
\centering
\includegraphics[width=\columnwidth]{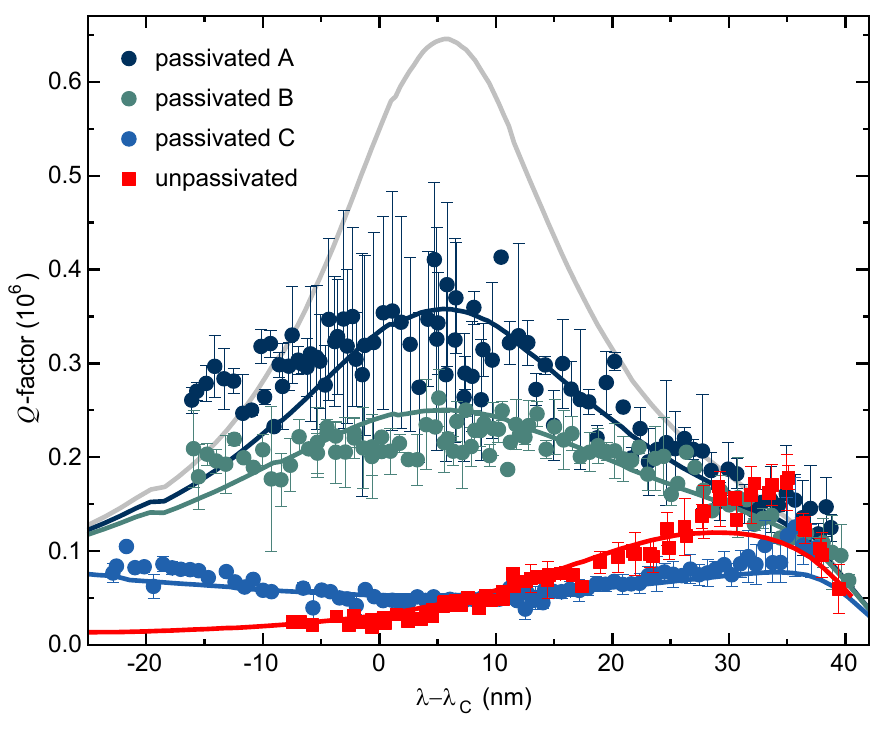}
\caption{Determination of surface-related losses via $\mathcal{Q}$-factor measurements using DBR-II with an nip-DBR. We probe the double role of surface passivation. Four positions in the contacted sample are probed: three passivated areas (circles) A, B, C and an unpassivated region (red squares). Mean values and standard deviation result from the measurements on the two cavity modes. We implement a mixed model to fit the data: F-K-like absorption and reasonable values of surface roughness are probed together. For the passivated regions, surface roughness alone explains the observed data. We estimate a roughness $\sigma = (\sigma_{\rm GaAs-Al2O3},\sigma_{\rm Al2O3-vac})$ for each region: at region A $\sigma=(0.5,0.5)$\,nm, alternatively $\sigma=(0.0,1.0)$\,nm; at region B $\sigma=(0.25,0.8)$\,nm, alternatively $\sigma=(0.0,1.4)$\,nm; at region C $\sigma=(0.5,3.0)$\,nm, alternatively $\sigma=(0.0,3.75)$\,nm. For the unpassivated region, surface roughness alone cannot account for the $\mathcal{Q}$-factor dependence on wavelength alone: with a surface roughness of $\sigma_{\rm GaAs-vac}=0.3$\,nm, an electric field $F_{\rm cap}=400$\,kV/cm is still needed to fit data.}
\label{cavitycharac_ECI}
\end{figure}

These results were verified in a second experiment employing a different piece of wafer material from the nip-DBR. The passivation was carried out in a separate run; the cavity was constructed with DBR-II as top mirror. The results are shown in Fig.~\ref{cavitycharac_ECI}. The results follow closely those of the first experiment, with the advantage that the $\mathcal{Q}$-factors could be determined also on the blue side of the SB center.

To quantify the loss, the entire microcavity is modeled (see Sec.~\ref{macleod}) using accurate descriptions of the two DBRs, including the free-carrier absorption in the doped layers in the heterostructure. Absorption is added to the capping layer (GaAs between the p-doping and the surface) and adjusted to match the experimentally determined $\mathcal{Q}$-factors at each wavelength. This is a robust procedure as the surface-related loss dominates other loss channels. The extracted absorption coefficients $\alpha$ from this procedure are plotted as a function of wavelength in Fig.~\ref{cavitycharac}(c). 

In both cases shown in Fig.~\ref{cavitycharac}(c) the absorption coefficients $\alpha$ are extremely small. From a measurement point of view, the microcavity represents a very sensitive platform for detecting very weak absorption or scattering events. In a single-pass experiment, these losses would be very difficult to detect. Significantly, we find that $\alpha$ depends exponentially on photon energy in the unpassivated case, a dependence that rules out scattering or broad-band absorption as the main loss mechanism at the surface, as these processes would have a much weaker dependence on wavelength. Instead, the exponential dependence points to below-gap absorption in an electric field.

\section{Investigation of loss via surface roughness}
\label{ucsb}

\begin{figure}[t!]
\centering
\includegraphics[width=\columnwidth]{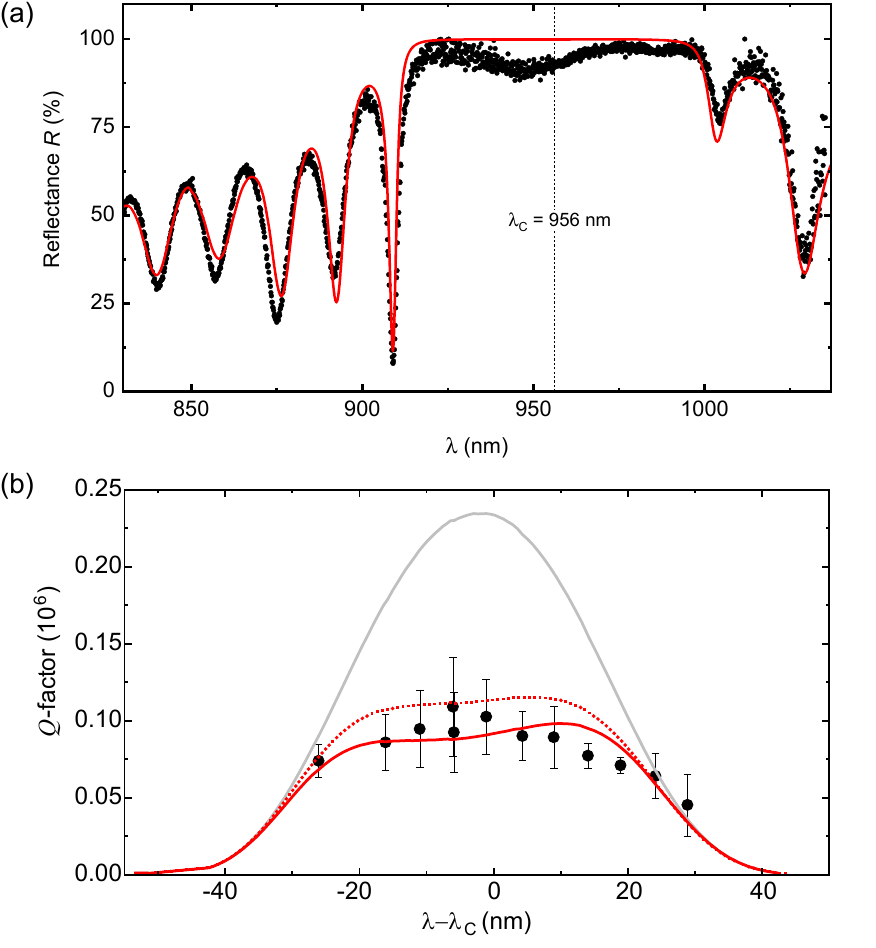}
\caption{Undoped semiconductor heterostructure: reflectance and $\mathcal{Q}$-factor measurements. (a) Reflectance measurement of a semiconductor heterostructure without doping ($\lambda_{\rm C}=956$\,nm) at $T=4.2$\,K. The heterostructure contains a $\lambda$-layer of GaAs on top of 33 pairs of AlAs($\lambda/4$)/GaAs($\lambda/4$). A layer of QDs is included at the center of the GaAs layer. In order to record a reference spectrum, parts of the sample were covered by an Au film using electron-beam evaporation. (b) $\mathcal{Q}$-factor of a microcavity at $T=4.2$\,K consisting of this heterostructure without doping paired with a dielectric top mirror DBR-I, experiment (points) and simulation with a surface roughness $\sigma_{\rm GaAs-vac}=0.5$\,nm (red solid line). The red dotted line is the expected $\mathcal{Q}$-factor on passivating the surface with a 8.0\,nm-thick Al$_2$O$_3$ layer. Gray solid line corresponds to expected $\mathcal{Q}$-factor without any surface losses.}
\label{a2}
\end{figure}

\begin{figure*}[t!]
\centering
\includegraphics[width=\textwidth]{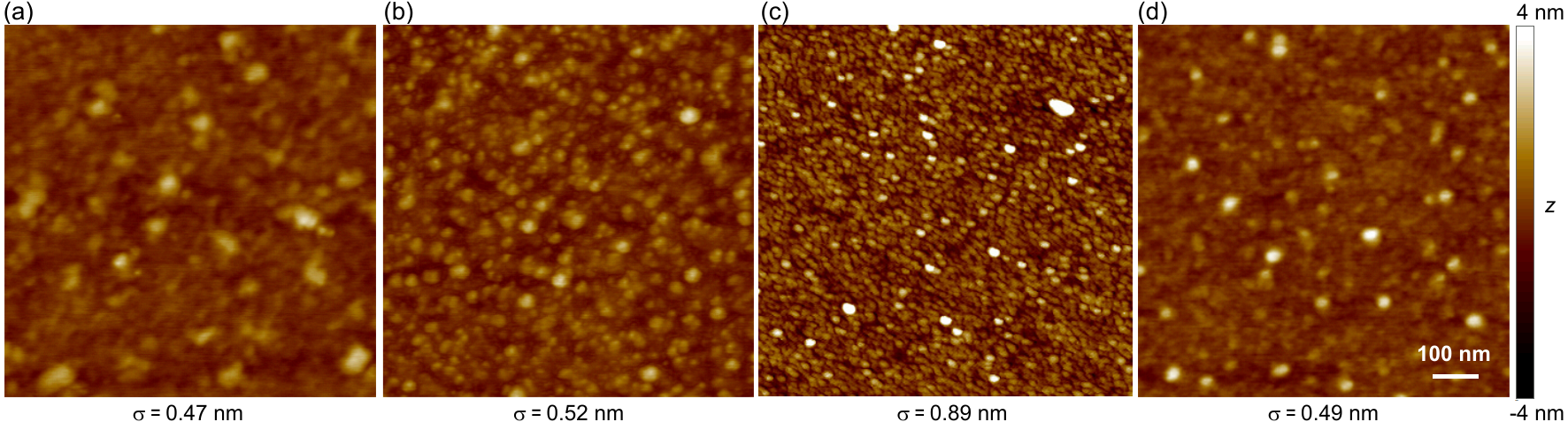}
\caption{Atomic force microscropy (AFM) images of different semiconductor samples. Each image depicts the same scan area of $0.8\times0.8\,\mu$m$^2$ (obtained in tapping mode) and scale bar from a height of $-4$ to $+4$\,nm. (a) Piece of passivated wafer. The rms surface roughness measured on several similar sample locations lies in the range $0.3\leq\sigma\leq1.9$\,nm. (b) Piece of unpassivated bare wafer. Range of surface roughness for similar sample locations is  $0.3\leq\sigma\leq0.7$\,nm. (c) Unpassivated sample with contacts. Similar samples present $0.4\leq\sigma\leq7.3$\,nm. (d) Piece of unpassivated bare wafer from a semiconductor DBR without n- and p-type layers. Different sample locations presented surface roughness ranging between $0.2\leq\sigma\leq0.9$\,nm.}
\label{fig:afm}
\end{figure*}

In order to confirm that the measured losses on the heterostructure with doping are indeed a consequence of the doped layers, and to quantify losses via surface scattering, we compare our results to a microcavity consisting of a semiconductor heterostructure without doping. The heterostructure in this case is a $\lambda$-layer of GaAs (with embedded InAs QDs in the center) on top of a 33-pair AlAs/GaAs DBR~\cite{Barbour2011,Greuter2015}\textsuperscript{,}\footnote{We note that the semiconductor heterostructure without doping was not grown with the same MBE as the semiconductor heterostructure with doping.} It is a high-quality sample but may not match the ultra-high quality of the semiconductor heterostructure with doping. Initially, we repeat the mirror characterization procedure described above and find a suitable model for the semiconductor layer thicknesses (Fig.~\ref{a2}(a)). Subsequently, we pair this mirror with the dielectric top mirror DBR-I.

Figure~\ref{a2}(b) depicts the measured $\mathcal{Q}$-factors as a function of wavelength for the semiconductor heterostructure without doping. The $\mathcal{Q}$-factor reaches approximately $10^{5}$ at the stopband center, and remains constant within measurement error over a wavelength range of approximately 20~nm. This behaviour is quite different to that of the unpassivated semiconductor heterostructure with doping (Fig.~\ref{cavitycharac}(b),(c), Fig.~\ref{cavitycharac_ECI}). The conclusion is that the strongly wavelength dependent loss process is related to the doping.

In the absence of losses, the heterostructure-without-doping -- top mirror combination should yield a $\mathcal{Q}$-factor of $2 \times 10^{5}$ in the stopband center, about a factor of two higher than that determined experimentally (Fig.~\ref{a2}(b), gray solid line). The wavelength-dependence of the $\mathcal{Q}$-factor (Fig.~\ref{a2}(b)) is, as before, a useful diagnostic of the scattering process. We find that in this case, scattering alone at the GaAs-vacuum interface can account for the measured $\mathcal{Q}$-factors, with a surface roughness $\sigma_{\rm GaAs-vac}=0.5$\,nm (Fig.~\ref{a2}(b) red solid line). We simulate as well the possible outcome of passivating the undoped sample, with $\sigma = (\sigma_{\rm GaAs-Al2O3},\sigma_{\rm Al2O3-vac})=(0.5,0.5)$\,nm, shown in Fig.~\ref{a2}(b) red dotted line: the addition of a thin alumina layer would diminish the effect of surface scattering due to surface roughness, increasing slightly the expected $\mathcal{Q}$-factor.

A surface roughness $\sigma$ translates into a total integrated scatter (TIS) of $\text{TIS} \approx(4\pi\sigma/\lambda)^2$ (Ref.~\cite{Bennett1992}) and can be modelled by an extinction coefficient $k$ (for the 1D transfer matrix methods) according to Ref.~\cite{Carniglia2002},
\begin{equation}
k=\frac{\pi(n_1-n_2)^2(n_1+n_2)d}{\lambda\sqrt{8(n_1^2+n_2^2)}},
\end{equation}
where $d=2\sigma$, $n_1$ and $n_2$ are the refractive indices of the two layers surrounding the scatter layer, and $\lambda$ is the free-space wavelength. Including this loss in the simulations reproduces the measured $\mathcal{Q}$-factors convincingly.
Furthermore, atomic force microscopy (AFM) measurements (tapping mode, Bruker Dimension 3100) indicate that surface roughness is present in different amounts across samples, as shown in Fig.~\ref{fig:afm}. The undoped wafer presented a root-mean-square (rms) surface roughness $0.3\leq\sigma\leq0.7$ \,nm, the unpassivated bare sample (without contacts) presented $0.2\leq\sigma\leq0.9$ \,nm, the unpassivated contacted sample $0.4\leq\sigma\leq7.3$ \,nm, and the passivated sample $0.3\leq\sigma\leq1.9$ \,nm. The native surface roughness at the GaAs-vacuum interface, right after growth of the samples, is on the order of 0.15 to 0.30\,nm. However, increased roughness is caused by processing of the samples -- e.g. via passivating, cleaning and gluing, bonding. We speculate that in the unpassivated contacted samples, presented high values of surface roughness might be induced by remnant traces of photoresist from the processing procedure (which in turn have a lowering scattering power than pure GaAs, due to the reduced refractive index).

The passivated doped sample also provided an opportunity to test the applicability of the TIS result for loss at a rough interface. The $\mathcal{Q}$-factors measured with DBR-II at three different passivated regions of the sample presented a weak dependence on wavelength, in contrast to the results in the unpassivated region. The model with $\sigma=(0.5,0.5)$\,nm, $\sigma=(0.25,0.8)$\,nm and $\sigma=(0.5,3.0)$\,nm respectively for regions A, B and C reproduce these $\mathcal{Q}$-factors precisely, with reasonable values of roughness at each interface (Fig.~\ref{cavitycharac_ECI}). Furthermore, the model captures the symmetric influence of roughness on the $\mathcal{Q}$-factor with respect to the SB center. This results from the reduction in the MCF at the sample's surface, as presented in Fig.~\ref{design2}(c). Agreement between the $\mathcal{Q}$-factors and the simulation gives us confidence that the description of loss via surface scattering is quantitatively correct.

The conclusion from this measurement and analysis is that, not only passivating the surface quenches surface-related absorption, but also has the beneficial aspect of reducing loss via surface scattering. The original GaAs surface and the passivated surface have a similar range of values of surface roughness. In the passivated case, the surface loss via scattering is about 60\% with respect to that of the unpassivated sample. The passivation procedure creates a layer with an intermediate refractive index: it avoids the large jump in refractive index at a GaAs-vacuum interface. This reduces the total scattering loss.

In the case of the doped unpassivated sample, the wavelength dependence of the $\mathcal{Q}$-factors is too strong to be accounted for with scattering loss (Fig.~\ref{cavitycharac}(b) and Fig.~\ref{cavitycharac_ECI}). A different mechanism is clearly at play.

\section{Microscopic explanation for the nip-DBR losses}

\begin{figure*}[t!]
\centering
\includegraphics[width=\textwidth]{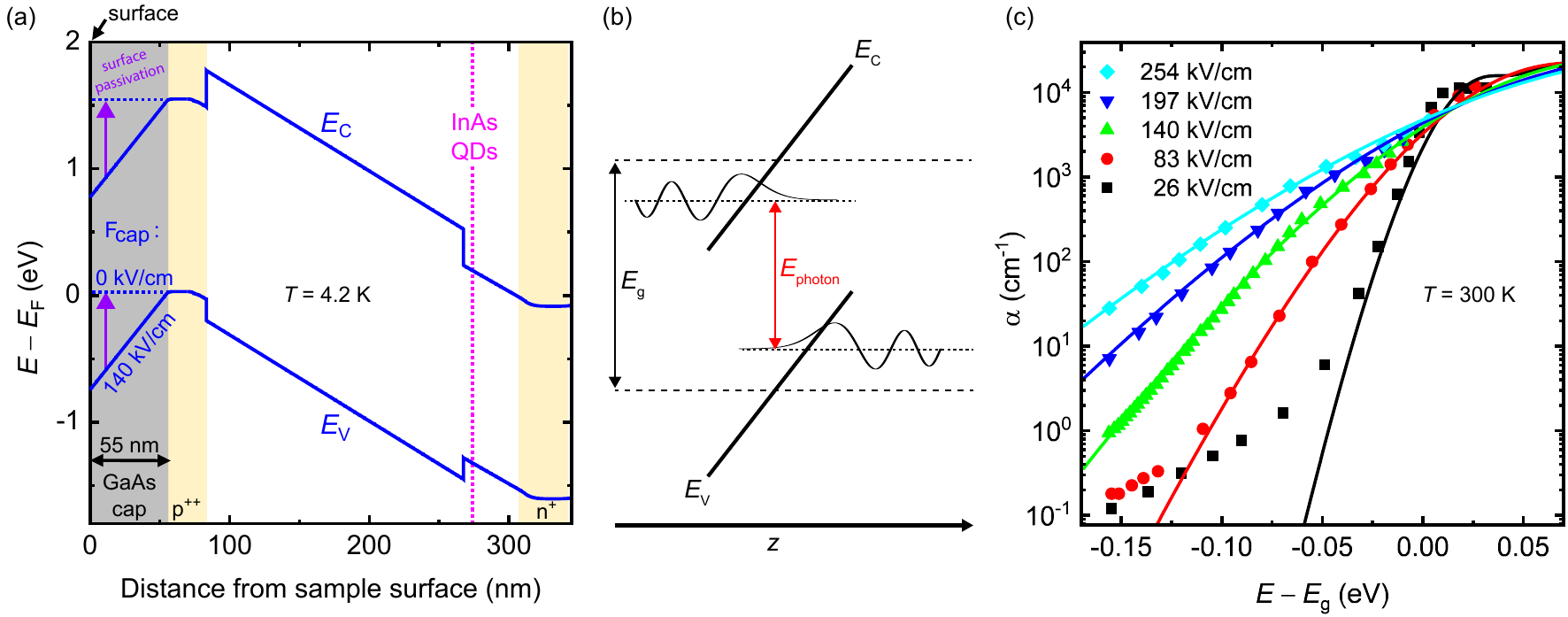}
\caption{Band structure and Franz-Keldysh effect. (a) Simulation of the conduction and valence bands in the n-i-p diode (nextnano) at $T=4.2$\,K. The surface is modelled via a Schottky barrier of $E_{\rm g}/2=0.76$\,eV reflecting the mid-gap Fermi-level pinning at the GaAs surface~\cite{Xuan2007}. The effect of surface passivation is to eliminate the electric field at the capping layer. (b) Schematic of the F-K effect~\cite{Franz1958,Keldysh1957}. An electric field applied to a semiconductor allows both electrons and holes to tunnel into the forbidden energy, leading to below-gap absorption processes. (c) Room-temperature F-K absorption coefficients for different electric fields within a p-i-n double heterostructure (\textcopyright 2020 IEEE. Reprinted, with permission, from Ref.~\cite{Knupfer1993}). The solid lines correspond to the calculated absorption coefficients according to Ref.~\cite{Aspnes1966,Hader1997} (Eqs.~\ref{4.1}--\ref{4.5}).}
\label{nextnano}
\end{figure*}

We give a possible microscopic explanation for the losses in the investigated nip-DBR structure and why surface passivation significantly reduces them. In Fig.~\ref{nextnano}(a), the calculated valence- and conduction-band edges in the heterostructure are shown, a solution to the 1D Poisson equation (obtained via the nextnano software). In the unpassivated case, we simulate the mid-gap Fermi-level pinning at the surface via a Schottky barrier of $0.76$\,eV. This yields an electric field in the capping layer (``capping field'') of $F_{\rm cap}=140$\,kV/cm. 

An electric field in a semiconductor leads to F-K absorption below the bandgap of the material~\cite{Franz1958,Keldysh1957}: owing to the position-dependence of the band-edges, the electron and hole wavefunctions can be described by Airy functions (similar to a particle in a triangular well~\cite{Davies1998}) and acquire an exponential tail at energies within the bandgap. The electric field therefore creates an absorption processes at photon energies $E_{\rm photon}<E_{\rm g}$. The situation is schematically depicted in Fig.~\ref{nextnano}(b).

According to the standard model~\cite{Aspnes1966}, F-K absorption at photon energy $E$ due to the presence of an electric field $F$ can be described via the absorption coefficient
\begin{multline}
\label{4.1}
\alpha(E,F)=\beta\cdot\frac{F^{1/3}}{E}\sum\limits_{i={\rm lh,hh}} \left(\frac{\mu_i}{m_0}\right)^{4/3} \\
\cdot|M_i|^2 \Big(|{\rm Ai}'(x_i)|^2-x_i|{\rm Ai}(x_i)|^2\Big),
\end{multline}
where
\begin{equation}
x_i=\frac{e\cdot\left(E_{\rm g}-E\right)}{\hbar\theta_i},
\end{equation}
\begin{equation}
\hbar\theta_i=\Big(\frac{(eF\hbar)^2}{2\mu_i}\Big)^{1/3}.
\end{equation}

\noindent In these equations, $\beta$ is a constant (arbitrary units), $e$ is the elementary charge (in SI units), $\hbar$ the reduced Planck's constant (in SI units), $m_0$ the free electron rest mass (in kg), $\mu_{\rm lh}=0.037m_0$ ($\mu_{\rm hh}=0.058m_0$) the reduced mass of an electron--light-hole pair (electron--heavy-hole pair), $F$ is given in kV/cm, the energies $E_g$ and $E$ in eV and $|M_{\rm lh}|^2$ ($|M_{\rm hh}|^2$) the momentum matrix elements for the light-hole (heavy-hole). ${\rm Ai}(z)$ is an Airy function\footnote{The Airy function is defined as ${\rm Ai}(z)=\frac{1}{2\pi}\int_{-\infty}^{\infty}e^{i(zt+t^3/3)}dt$.} (with derivative ${\rm Ai}'(z)$).

We make use of the momentum matrix elements derived in Ref.~\cite{Hader1997} for different polarizations of the radiation field. For light polarized in the $(x,y)$-plane, the momentum matrix elements for the light- and heavy-holes read
\begin{align}
|M_{\rm lh}|^2&=P^2/3,\\
|M_{\rm hh}|^2&=P^2,
\label{4.5}
\end{align}
where $P=0.692$\,(arbitrary units) is a typical value for GaAs~\cite{Hader1997}.

We use this model for F-K absorption to describe previously reported room-temperature experiments on a p-i-n double heterostructure~\cite{Knupfer1993} (Fig.~\ref{nextnano}(c)), and extract the value of constant $\beta$ in Eq.~\ref{4.1}, which is found to be $\beta = 2.5 \cdot 10^4$\,(arbitrary units). These experiments extend to photon energies far below the bandgap, the case of interest here. There is a compelling overlap between theory and experiment. 

In order to estimate F-K absorption coefficients in our nip-DBR at 4.2\,K (Fig.~\ref{cavitycharac}(b),(c) and Fig.~\ref{cavitycharac_ECI}), we make use of Eqs.~\ref{4.1}--\ref{4.5}, taking the low-temperature GaAs bandgap of 1.519\,eV and $\beta$ extracted from fitting the data in Fig.~\ref{nextnano}(c). We compare the results of the model for low-temperature F-K absorption to the experimental data presented in Fig.~\ref{cavitycharac}(b),(c), taking the electric field $F$ as a fitting parameter. The exponential dependence of the absorption on the photon energy is well described with F-K absorption. However, in the unpassivated case, the capping field is $F_{\rm cap}=470$\,kV/cm when not accounting for any surface roughness (Fig.~\ref{cavitycharac}) and $F_{\rm cap}=400$\,kV/cm when accounting for a realistic value of surface roughness (Fig.~\ref{cavitycharac_ECI}). These values are between 3.4 and 2.8 above that expected from the 1D Poisson equation (Fig.~\ref{nextnano}(a)) respectively. The origin of this discrepancy is not understood at this point, but we note several points.

First, to the best of our knowledge, there are no F-K absorption experiments reported in the literature at low temperature (4.2\,K) and at photon energies far below the bandgap $E_{\rm g}$ of GaAs (at $E-E_{\rm g}\sim-0.17$\,eV corresponding to $\lambda\sim920$\,nm). Our approach here is to fit the theory presented in Ref.~\cite{Aspnes1966,Hader1997} to the room temperature experiments of Ref.~\cite{Knupfer1993} (Fig.~\ref{nextnano}(c)) and then to extrapolate the absorption coefficients to photon energies $\sim 0.17$\,eV below the bandgap. The change in temperature, from room-temperate to low temperature, is accommodated by a rigid shift in the absorption spectrum to account for the increase in the bandgap. It is conceivable that the standard F-K theory is inadequate at photon energies far below the bandgap -- this point has not been tested experimentally.

Secondly, there are room-temperature experiments on F-K oscillations in doped GaAs heterostructures (a 25--80\,nm thick, undoped GaAs capping layer on top of an n$^+$-doped Al$_{.32}$Ga$_{.68}$As layer~\cite{Zamora2016}) that also report surface electric field values a factor 1.8--3.8 above the expected ones\footnote{In order to estimate the capping fields in Ref.~\cite{Zamora2016}, we assume mid-gap pinning, dividing half the bandgap $E_{\rm g}/2=0.71$\,eV of GaAs at 300\,K by the capping layer thicknesses there reported.}.

Thirdly, the F-K model describes a bulk semiconductor in a uniform electric field. Obviously, it does not take into account the microscopic details of the surface, for instance surface reconstructions and oxidation. It is possible that the details of the surface layer are important here.

In the light of this analysis, our proposal is that surface passivation quenches surface-related absorption primarily by reducing the electric field in the capping layer, thereby eliminating the F-K absorption (within the sensitivity of the experiment).

A remaining question is why the bare-wafer sample without passivation shows higher $\mathcal{Q}$-factors than the electrically contacted sample also without passivation (Fig.~\ref{cavitycharac}(b),(c)). The contacting process may change the surface roughness, as revealed by the AFM measurements (Fig.~\ref{fig:afm}). The surface roughness per se does not however account for the $\mathcal{Q}$-factors of the various samples. The main point is that surface scattering does not account for the exponential dependence of the loss process on photon energy (Fig.~\ref{cavitycharac}(b) and (c)). Instead, we speculate that the change in GaAs surface on forming the contacts results in a change of surface pinning, thereby increasing the capping field. One possibility is that the degradation of the surface on contacting spreads the available surface states to lower energies.

\section{Model for the curved dielectric mirrors}

\begin{figure}[t!]
\centering
\includegraphics[width=\columnwidth]{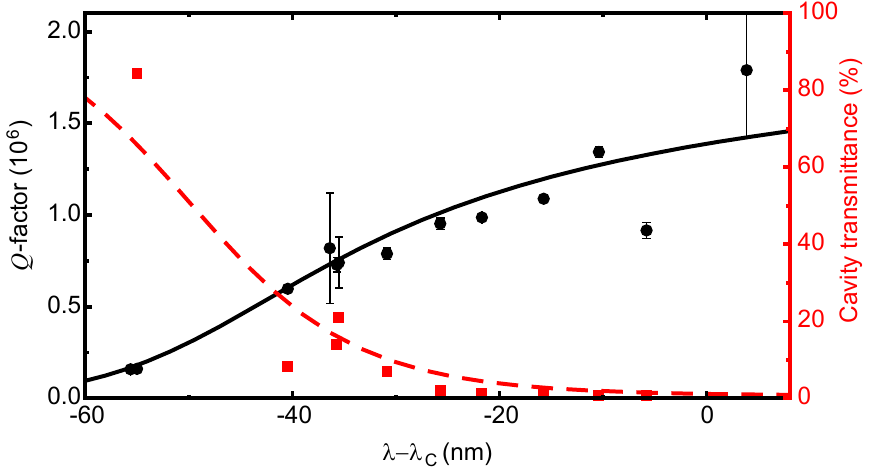}
\caption{Measured $\mathcal{Q}$-factors and cavity transmittance of a purely dielectric microcavity. Dielectric top mirror ($\lambda_{\rm C}=973$\,nm) paired with a dielectric bottom mirror (shifted to $\lambda_{\rm C}=976$\,nm, nominally the same coating) at $T=300$\,K. The cavity transmittance is measured by relating the transmitted power at the cavity resonance to the laser power before the objective lens multiplied by a fitted in-coupling efficiency of 59\%. The black solid (red dashed) line is a calculation of the $\mathcal{Q}$-factor (cavity transmittance) taking into account a material extinction coefficient of $k_{\rm SiO2}=4\cdot10^{-7}$ and $k_{\rm Ta2O5}=4.5\cdot10^{-7}$ for SiO$_2$ and Ta$_2$O$_5$, respectively~\cite{BeauvilleCQG2004}. Additionally, an interface roughness of $\sigma=0.25$\,nm above each of the five last grown Ta$_2$O$_5$ layers and $k=4k_{\rm Ta2O5}$ in the last-grown Ta$_2$O$_5$ layer are introduced heuristically in order to fit the experimental data.}
\label{a1}
\end{figure}

An interpretation of the microcavity $\mathcal{Q}$-factors in terms of losses in the semiconductor heterostructure rests on an understanding of the top mirror. The top mirrors, dielectric-DBRs, are of very high quality with very low loss. To prove this point, we investigate a microcavity formed from DBR-I and a planar version of DBR-I. The coatings for the plane mirror and concave mirrors were applied to the substrates in the same run and are nominally identical. Fig.~\ref{a1} shows the measured $\mathcal{Q}$-factors. At the stopband center of the top mirror ($\lambda_{\rm C}=976$\,nm), the $\mathcal{Q}$-factor is extremely high, $1.5 \cdot 10^{6}$. To describe the dielectric mirror accurately at the stopband center of the semiconductor-DBR, we analyze the dependence of the dielectric-dielectric $\mathcal{Q}$-factor and transmission as a function of wavelength. To describe the high transmission at short wavelengths, we are forced to red-shift the stopband center of the bottom mirror by 3\,nm\footnote{We note that reflectance spectra (Fig.~\ref{mirrorcharac}(a)) on different samples with nominally the same dielectric coating exhibited up to 6\,nm shifts in wavelength, most probably due to thickness variations across the wafer.}. A rough interface at the ``lower" surface of the five ``lowest" Ta$_2$O$_5$ layers shown in Fig.~\ref{design2} (extinction coefficients corresponding to an interface roughness~\cite{Carniglia2002} of 0.25 nm) together with an increased absorption within the ``lowest" Ta$_2$O$_5$ layer (extinction coefficient $k=4k_{\rm Ta2O5}$, where $k_{\rm Ta2O5}$ is defined in Sec.~\ref{macleod}) are heuristically introduced in the model in order to describe the measured $\mathcal{Q}$-factors. This fit is very convincing (Fig.~\ref{a1}). This description of the top dielectric-mirror DBR-I is used to interpret the measurements on microcavities formed using the semiconductor-DBRs as bottom mirror (Fig.~\ref{cavitycharac}(b),(c) and Fig.~\ref{a2}(b)).

\section{1D transfer matrix calculation of the microcavity $\mathcal{Q}$-factors}
\label{macleod}
We list the relevant parameters used for the $\mathcal{Q}$-factor calculations via a one-dimensional transfer-matrix method.

Extinction coefficients in the different materials are introduced as follows. For DBR-I, $k_{\rm SiO2}=4\cdot10^{-7}$ for SiO$_2$ (Ref.~\cite{BeauvilleCQG2004}), $k_{\rm Ta2O5}=4.5\cdot10^{-7}$ for Ta$_2$O$_5$ (Ref.~\cite{BeauvilleCQG2004}), and $k=4k_{\rm Ta2O5}$ is used for the ``lowest", i.e.\ the last-grown, Ta$_2$O$_5$ layer (the layer closest to the vacuum-gap). For DBR-II, $k_{\rm SiO2}=4\cdot10^{-6}$ for SiO$_2$, $k_{\rm Ta2O5}=4.5\cdot10^{-6}$ for Ta$_2$O$_5$ (from fitting data taken with curved DBR-II and planar DBR-I microcavity). Finally, for the semiconductor nip-DBR, $k_{\rm p++}=5.2\cdot10^{-4}$ for p$^{++}$-GaAs, $k_{\rm p+}=1.9\cdot10^{-4}$ for p$^{+}$-GaAs, and $k_{\rm n+}=0.7\cdot10^{-4}$ for n$^{+}$-GaAs~\cite{Casey1975}. An extinction coefficient of $k=\alpha_{\rm FK}\lambda/(4\pi)$, where $\alpha_{\rm FK}$ is a F-K absorption coefficient, is introduced in the capping layer; $\alpha_{\rm FK}$ depends on the electric field. Surface roughness is described by introducing an additional layer of thickness $d=2\sigma$, where $\sigma$ is the rms surface/interface roughness~\cite{Carniglia2002}.

In analogy to the experiment, a $\mathcal{Q}$-factor is determined for a fixed ``vacuum-gap'' layer thickness by calculating a cavity transmittance spectrum. A Lorentzian fit to the calculated spectrum is used to determine the resonance frequency as well as the $\mathcal{Q}$-factor. This procedure is repeated for different vacuum-gaps, yielding a plot of $\mathcal{Q}$-factor versus wavelength. The resulting $\mathcal{Q}$-factors are presented in Fig.~\ref{cavitycharac}, Fig.~\ref{cavitycharac_ECI}, Fig.~\ref{a2} and Fig.~\ref{a1}.

\section{Conclusion}
\label{conclusion}
Significant surface-related losses in an open microcavity consisting of an nip-DBR and dielectric DBR are much reduced by passivating the GaAs surface. The passivation works primarily by eliminating the Franz-Keldysh-like absorption in the capping layer. Secondary benefit of the passivation is to reduce loss by surface scattering. With passivation, $\mathcal{Q}$-factors close to $10^6$ can be achieved.

\section*{Acknowledgements}
We thank Ivan Favero for inspiration; we thank Leonardo Midolo and Peter Lodahl for fruitful discussions. We thank Monica Sch\"onenberger and the Nano Imaging Lab (University of Basel) for support acquiring AFM images. This work was funded by Swiss National Science Foundation Project 200020\_175748, NCCR QSIT and European Union Horizon2020 FET-Open Project QLUSTER. A.J.\ acknowledges support from the European Union Horizon 2020 Research and Innovation Programme under the Marie Sk{\l}odowska-Curie grant agreement No.\ 840453 (HiFig). S.R.V., R.S., A.L.\ and A.D.W.\ acknowledge gratefully support from DFH/UFA CDFA05-06, DFG TRR160, DFG project 383065199, and BMBF Q.Link.X.\\

\bibliography{Najer_SurfacePassivation_2020_v4}

\begin{thebibliography}{39}%
\makeatletter
\providecommand \@ifxundefined [1]{%
 \@ifx{#1\undefined}
}%
\providecommand \@ifnum [1]{%
 \ifnum #1\expandafter \@firstoftwo
 \else \expandafter \@secondoftwo
 \fi
}%
\providecommand \@ifx [1]{%
 \ifx #1\expandafter \@firstoftwo
 \else \expandafter \@secondoftwo
 \fi
}%
\providecommand \natexlab [1]{#1}%
\providecommand \enquote  [1]{``#1''}%
\providecommand \bibnamefont  [1]{#1}%
\providecommand \bibfnamefont [1]{#1}%
\providecommand \citenamefont [1]{#1}%
\providecommand \href@noop [0]{\@secondoftwo}%
\providecommand \href [0]{\begingroup \@sanitize@url \@href}%
\providecommand \@href[1]{\@@startlink{#1}\@@href}%
\providecommand \@@href[1]{\endgroup#1\@@endlink}%
\providecommand \@sanitize@url [0]{\catcode `\\12\catcode `\$12\catcode
  `\&12\catcode `\#12\catcode `\^12\catcode `\_12\catcode `\%12\relax}%
\providecommand \@@startlink[1]{}%
\providecommand \@@endlink[0]{}%
\providecommand \url  [0]{\begingroup\@sanitize@url \@url }%
\providecommand \@url [1]{\endgroup\@href {#1}{\urlprefix }}%
\providecommand \urlprefix  [0]{URL }%
\providecommand \Eprint [0]{\href }%
\providecommand \doibase [0]{http://dx.doi.org/}%
\providecommand \selectlanguage [0]{\@gobble}%
\providecommand \bibinfo  [0]{\@secondoftwo}%
\providecommand \bibfield  [0]{\@secondoftwo}%
\providecommand \translation [1]{[#1]}%
\providecommand \BibitemOpen [0]{}%
\providecommand \bibitemStop [0]{}%
\providecommand \bibitemNoStop [0]{.\EOS\space}%
\providecommand \EOS [0]{\spacefactor3000\relax}%
\providecommand \BibitemShut  [1]{\csname bibitem#1\endcsname}%
\let\auto@bib@innerbib\@empty
\bibitem [{\citenamefont {Reithmaier}\ \emph {et~al.}(2004)\citenamefont
  {Reithmaier}, \citenamefont {Sek}, \citenamefont {L\"offler}, \citenamefont
  {Hofmann}, \citenamefont {Kuhn}, \citenamefont {Reitzenstein}, \citenamefont
  {Keldysh}, \citenamefont {Kulakovskii}, \citenamefont {Reinecke},\ and\
  \citenamefont {Forchel}}]{Reithmaier2004}%
  \BibitemOpen
  \bibfield  {author} {\bibinfo {author} {\bibfnamefont {J.}~\bibnamefont
  {Reithmaier}}, \bibinfo {author} {\bibfnamefont {G.}~\bibnamefont {Sek}},
  \bibinfo {author} {\bibfnamefont {A.}~\bibnamefont {L\"offler}}, \bibinfo
  {author} {\bibfnamefont {C.}~\bibnamefont {Hofmann}}, \bibinfo {author}
  {\bibfnamefont {S.}~\bibnamefont {Kuhn}}, \bibinfo {author} {\bibfnamefont
  {S.}~\bibnamefont {Reitzenstein}}, \bibinfo {author} {\bibfnamefont
  {L.}~\bibnamefont {Keldysh}}, \bibinfo {author} {\bibfnamefont
  {V.}~\bibnamefont {Kulakovskii}}, \bibinfo {author} {\bibfnamefont
  {T.}~\bibnamefont {Reinecke}}, \ and\ \bibinfo {author} {\bibfnamefont
  {A.}~\bibnamefont {Forchel}},\ }\bibfield  {title} {\enquote {\bibinfo
  {title} {Strong coupling in a single quantum dot-semiconductor microcavity
  system},}\ }\href {\doibase 10.1038/nature02969} {\bibfield  {journal}
  {\bibinfo  {journal} {Nature}\ }\textbf {\bibinfo {volume} {432}},\ \bibinfo
  {pages} {197--200} (\bibinfo {year} {2004})}\BibitemShut {NoStop}%
\bibitem [{\citenamefont {Somaschi}\ \emph {et~al.}(2016)\citenamefont
  {Somaschi}, \citenamefont {Giesz}, \citenamefont {De~Santis}, \citenamefont
  {Loredo}, \citenamefont {Almeida}, \citenamefont {Hornecker}, \citenamefont
  {Portalupi}, \citenamefont {Grange}, \citenamefont {Anton}, \citenamefont
  {Demory}, \citenamefont {Gomez}, \citenamefont {Sagnes}, \citenamefont
  {Lanzillotti-Kimura}, \citenamefont {Lemaitre}, \citenamefont {Auffeves},
  \citenamefont {White}, \citenamefont {Lanco},\ and\ \citenamefont
  {Senellart}}]{Somaschi2016}%
  \BibitemOpen
  \bibfield  {author} {\bibinfo {author} {\bibfnamefont {N.}~\bibnamefont
  {Somaschi}}, \bibinfo {author} {\bibfnamefont {V.}~\bibnamefont {Giesz}},
  \bibinfo {author} {\bibfnamefont {L.}~\bibnamefont {De~Santis}}, \bibinfo
  {author} {\bibfnamefont {J.~C.}\ \bibnamefont {Loredo}}, \bibinfo {author}
  {\bibfnamefont {M.~P.}\ \bibnamefont {Almeida}}, \bibinfo {author}
  {\bibfnamefont {G.}~\bibnamefont {Hornecker}}, \bibinfo {author}
  {\bibfnamefont {S.~L.}\ \bibnamefont {Portalupi}}, \bibinfo {author}
  {\bibfnamefont {T.}~\bibnamefont {Grange}}, \bibinfo {author} {\bibfnamefont
  {C.}~\bibnamefont {Anton}}, \bibinfo {author} {\bibfnamefont
  {J.}~\bibnamefont {Demory}}, \bibinfo {author} {\bibfnamefont
  {C.}~\bibnamefont {Gomez}}, \bibinfo {author} {\bibfnamefont
  {I.}~\bibnamefont {Sagnes}}, \bibinfo {author} {\bibfnamefont {N.~D.}\
  \bibnamefont {Lanzillotti-Kimura}}, \bibinfo {author} {\bibfnamefont
  {A.}~\bibnamefont {Lemaitre}}, \bibinfo {author} {\bibfnamefont
  {A.}~\bibnamefont {Auffeves}}, \bibinfo {author} {\bibfnamefont {A.~G.}\
  \bibnamefont {White}}, \bibinfo {author} {\bibfnamefont {L.}~\bibnamefont
  {Lanco}}, \ and\ \bibinfo {author} {\bibfnamefont {P.}~\bibnamefont
  {Senellart}},\ }\bibfield  {title} {\enquote {\bibinfo {title} {Near-optimal
  single-photon sources in the solid state},}\ }\href {\doibase
  10.1038/NPHOTON.2016.23} {\bibfield  {journal} {\bibinfo  {journal} {Nat.
  Photon.}\ }\textbf {\bibinfo {volume} {10}},\ \bibinfo {pages} {340--345}
  (\bibinfo {year} {2016})}\BibitemShut {NoStop}%
\bibitem [{\citenamefont {Ding}\ \emph {et~al.}(2016)\citenamefont {Ding},
  \citenamefont {He}, \citenamefont {Duan}, \citenamefont {Gregersen},
  \citenamefont {Chen}, \citenamefont {Unsleber}, \citenamefont {Maier},
  \citenamefont {Schneider}, \citenamefont {Kamp}, \citenamefont {H\"ofling},
  \citenamefont {Lu},\ and\ \citenamefont {Pan}}]{Ding2016}%
  \BibitemOpen
  \bibfield  {author} {\bibinfo {author} {\bibfnamefont {X.}~\bibnamefont
  {Ding}}, \bibinfo {author} {\bibfnamefont {Y.}~\bibnamefont {He}}, \bibinfo
  {author} {\bibfnamefont {Z.-C.}\ \bibnamefont {Duan}}, \bibinfo {author}
  {\bibfnamefont {N.}~\bibnamefont {Gregersen}}, \bibinfo {author}
  {\bibfnamefont {M.-C.}\ \bibnamefont {Chen}}, \bibinfo {author}
  {\bibfnamefont {S.}~\bibnamefont {Unsleber}}, \bibinfo {author}
  {\bibfnamefont {S.}~\bibnamefont {Maier}}, \bibinfo {author} {\bibfnamefont
  {C.}~\bibnamefont {Schneider}}, \bibinfo {author} {\bibfnamefont
  {M.}~\bibnamefont {Kamp}}, \bibinfo {author} {\bibfnamefont {S.}~\bibnamefont
  {H\"ofling}}, \bibinfo {author} {\bibfnamefont {C.-Y.}\ \bibnamefont {Lu}}, \
  and\ \bibinfo {author} {\bibfnamefont {J.-W.}\ \bibnamefont {Pan}},\
  }\bibfield  {title} {\enquote {\bibinfo {title} {On-demand single photons
  with high extraction efficiency and near-unity indistinguishability from a
  resonantly driven quantum dot in a micropillar},}\ }\href {\doibase
  10.1103/PhysRevLett.116.020401} {\bibfield  {journal} {\bibinfo  {journal}
  {Phys. Rev. Lett.}\ }\textbf {\bibinfo {volume} {116}},\ \bibinfo {pages}
  {020401} (\bibinfo {year} {2016})}\BibitemShut {NoStop}%
\bibitem [{\citenamefont {Yoshie}\ \emph {et~al.}(2004)\citenamefont {Yoshie},
  \citenamefont {Scherer}, \citenamefont {Hendrickson}, \citenamefont
  {Khitrova}, \citenamefont {Gibbs}, \citenamefont {Rupper}, \citenamefont
  {Ell}, \citenamefont {Shchekin},\ and\ \citenamefont {Deppe}}]{Yoshie2004}%
  \BibitemOpen
  \bibfield  {author} {\bibinfo {author} {\bibfnamefont {T.}~\bibnamefont
  {Yoshie}}, \bibinfo {author} {\bibfnamefont {A.}~\bibnamefont {Scherer}},
  \bibinfo {author} {\bibfnamefont {J.}~\bibnamefont {Hendrickson}}, \bibinfo
  {author} {\bibfnamefont {G.}~\bibnamefont {Khitrova}}, \bibinfo {author}
  {\bibfnamefont {H.}~\bibnamefont {Gibbs}}, \bibinfo {author} {\bibfnamefont
  {G.}~\bibnamefont {Rupper}}, \bibinfo {author} {\bibfnamefont
  {C.}~\bibnamefont {Ell}}, \bibinfo {author} {\bibfnamefont {O.}~\bibnamefont
  {Shchekin}}, \ and\ \bibinfo {author} {\bibfnamefont {D.}~\bibnamefont
  {Deppe}},\ }\bibfield  {title} {\enquote {\bibinfo {title} {Vacuum {R}abi
  splitting with a single quantum dot in a photonic crystal nanocavity},}\
  }\href {\doibase 10.1038/nature03119} {\bibfield  {journal} {\bibinfo
  {journal} {Nature}\ }\textbf {\bibinfo {volume} {432}},\ \bibinfo {pages}
  {200--203} (\bibinfo {year} {2004})}\BibitemShut {NoStop}%
\bibitem [{\citenamefont {Kuruma}\ \emph {et~al.}(2020)\citenamefont {Kuruma},
  \citenamefont {Ota}, \citenamefont {Kakuda}, \citenamefont {Iwamoto},\ and\
  \citenamefont {Arakawa}}]{Kuruma2020}%
  \BibitemOpen
  \bibfield  {author} {\bibinfo {author} {\bibfnamefont {K.}~\bibnamefont
  {Kuruma}}, \bibinfo {author} {\bibfnamefont {Y.}~\bibnamefont {Ota}},
  \bibinfo {author} {\bibfnamefont {M.}~\bibnamefont {Kakuda}}, \bibinfo
  {author} {\bibfnamefont {S.}~\bibnamefont {Iwamoto}}, \ and\ \bibinfo
  {author} {\bibfnamefont {Y.}~\bibnamefont {Arakawa}},\ }\bibfield  {title}
  {\enquote {\bibinfo {title} {Surface-passivated high-q gaas photonic crystal
  nanocavity with quantum dots},}\ }\href {\doibase 10.1063/1.5144959}
  {\bibfield  {journal} {\bibinfo  {journal} {APL Photonics}\ }\textbf
  {\bibinfo {volume} {5}},\ \bibinfo {pages} {046106} (\bibinfo {year}
  {2020})}\BibitemShut {NoStop}%
\bibitem [{\citenamefont {Guha}\ \emph {et~al.}(2017)\citenamefont {Guha},
  \citenamefont {Marsault}, \citenamefont {Cadiz}, \citenamefont {Morgenroth},
  \citenamefont {Ulin}, \citenamefont {Berkovitz}, \citenamefont {Lemaitre},
  \citenamefont {Gomez}, \citenamefont {Amo}, \citenamefont {Combrie},
  \citenamefont {Gerard}, \citenamefont {Leo},\ and\ \citenamefont
  {Favero}}]{Guha2017}%
  \BibitemOpen
  \bibfield  {author} {\bibinfo {author} {\bibfnamefont {B.}~\bibnamefont
  {Guha}}, \bibinfo {author} {\bibfnamefont {F.}~\bibnamefont {Marsault}},
  \bibinfo {author} {\bibfnamefont {F.}~\bibnamefont {Cadiz}}, \bibinfo
  {author} {\bibfnamefont {L.}~\bibnamefont {Morgenroth}}, \bibinfo {author}
  {\bibfnamefont {V.}~\bibnamefont {Ulin}}, \bibinfo {author} {\bibfnamefont
  {V.}~\bibnamefont {Berkovitz}}, \bibinfo {author} {\bibfnamefont
  {A.}~\bibnamefont {Lemaitre}}, \bibinfo {author} {\bibfnamefont
  {C.}~\bibnamefont {Gomez}}, \bibinfo {author} {\bibfnamefont
  {A.}~\bibnamefont {Amo}}, \bibinfo {author} {\bibfnamefont {S.}~\bibnamefont
  {Combrie}}, \bibinfo {author} {\bibfnamefont {B.}~\bibnamefont {Gerard}},
  \bibinfo {author} {\bibfnamefont {G.}~\bibnamefont {Leo}}, \ and\ \bibinfo
  {author} {\bibfnamefont {I.}~\bibnamefont {Favero}},\ }\bibfield  {title}
  {\enquote {\bibinfo {title} {Surface-enhanced gallium arsenide photonic
  resonator with quality factor of $6 \times 10^{6}$},}\ }\href {\doibase
  10.1364/OPTICA.4.000218} {\bibfield  {journal} {\bibinfo  {journal} {Optica}\
  }\textbf {\bibinfo {volume} {4}},\ \bibinfo {pages} {218--221} (\bibinfo
  {year} {2017})}\BibitemShut {NoStop}%
\bibitem [{\citenamefont {Barbour}\ \emph {et~al.}(2011)\citenamefont
  {Barbour}, \citenamefont {Dalgarno}, \citenamefont {Curran}, \citenamefont
  {Nowak}, \citenamefont {Baker}, \citenamefont {Hall}, \citenamefont {Stoltz},
  \citenamefont {Petroff},\ and\ \citenamefont {Warburton}}]{Barbour2011}%
  \BibitemOpen
  \bibfield  {author} {\bibinfo {author} {\bibfnamefont {R.~J.}\ \bibnamefont
  {Barbour}}, \bibinfo {author} {\bibfnamefont {P.~A.}\ \bibnamefont
  {Dalgarno}}, \bibinfo {author} {\bibfnamefont {A.}~\bibnamefont {Curran}},
  \bibinfo {author} {\bibfnamefont {K.~M.}\ \bibnamefont {Nowak}}, \bibinfo
  {author} {\bibfnamefont {H.~J.}\ \bibnamefont {Baker}}, \bibinfo {author}
  {\bibfnamefont {D.~R.}\ \bibnamefont {Hall}}, \bibinfo {author}
  {\bibfnamefont {N.~G.}\ \bibnamefont {Stoltz}}, \bibinfo {author}
  {\bibfnamefont {P.~M.}\ \bibnamefont {Petroff}}, \ and\ \bibinfo {author}
  {\bibfnamefont {R.~J.}\ \bibnamefont {Warburton}},\ }\bibfield  {title}
  {\enquote {\bibinfo {title} {A tunable microcavity},}\ }\href@noop {}
  {\bibfield  {journal} {\bibinfo  {journal} {J. Appl. Phys.}\ }\textbf
  {\bibinfo {volume} {110}},\ \bibinfo {pages} {053107} (\bibinfo {year}
  {2011})}\BibitemShut {NoStop}%
\bibitem [{\citenamefont {Greuter}\ \emph {et~al.}(2015)\citenamefont
  {Greuter}, \citenamefont {Starosielec}, \citenamefont {Kuhlmann},\ and\
  \citenamefont {Warburton}}]{Greuter2015}%
  \BibitemOpen
  \bibfield  {author} {\bibinfo {author} {\bibfnamefont {L.}~\bibnamefont
  {Greuter}}, \bibinfo {author} {\bibfnamefont {S.}~\bibnamefont
  {Starosielec}}, \bibinfo {author} {\bibfnamefont {A.~V.}\ \bibnamefont
  {Kuhlmann}}, \ and\ \bibinfo {author} {\bibfnamefont {R.~J.}\ \bibnamefont
  {Warburton}},\ }\bibfield  {title} {\enquote {\bibinfo {title} {Towards
  high-cooperativity strong coupling of a quantum dot in a tunable
  microcavity},}\ }\href {\doibase 10.1103/PhysRevB.92.045302} {\bibfield
  {journal} {\bibinfo  {journal} {Phys. Rev. B}\ }\textbf {\bibinfo {volume}
  {92}},\ \bibinfo {pages} {045302} (\bibinfo {year} {2015})}\BibitemShut
  {NoStop}%
\bibitem [{\citenamefont {Najer}\ \emph {et~al.}(2019)\citenamefont {Najer},
  \citenamefont {S\"{o}llner}, \citenamefont {Sekatski}, \citenamefont
  {Dolique}, \citenamefont {L\"{o}bl}, \citenamefont {Riedel}, \citenamefont
  {Schott}, \citenamefont {Starosielec}, \citenamefont {Valentin},
  \citenamefont {Wieck}, \citenamefont {Sangouard}, \citenamefont {Ludwig},\
  and\ \citenamefont {Warburton}}]{Najer2019}%
  \BibitemOpen
  \bibfield  {author} {\bibinfo {author} {\bibfnamefont {D.}~\bibnamefont
  {Najer}}, \bibinfo {author} {\bibfnamefont {I.}~\bibnamefont {S\"{o}llner}},
  \bibinfo {author} {\bibfnamefont {P.}~\bibnamefont {Sekatski}}, \bibinfo
  {author} {\bibfnamefont {V.}~\bibnamefont {Dolique}}, \bibinfo {author}
  {\bibfnamefont {M.~C.}\ \bibnamefont {L\"{o}bl}}, \bibinfo {author}
  {\bibfnamefont {D.}~\bibnamefont {Riedel}}, \bibinfo {author} {\bibfnamefont
  {R.}~\bibnamefont {Schott}}, \bibinfo {author} {\bibfnamefont
  {S.}~\bibnamefont {Starosielec}}, \bibinfo {author} {\bibfnamefont {S.~R.}\
  \bibnamefont {Valentin}}, \bibinfo {author} {\bibfnamefont {A.~D.}\
  \bibnamefont {Wieck}}, \bibinfo {author} {\bibfnamefont {N.}~\bibnamefont
  {Sangouard}}, \bibinfo {author} {\bibfnamefont {A.}~\bibnamefont {Ludwig}}, \
  and\ \bibinfo {author} {\bibfnamefont {R.~J.}\ \bibnamefont {Warburton}},\
  }\bibfield  {title} {\enquote {\bibinfo {title} {A gated quantum dot strongly
  coupled to an optical microcavity},}\ }\href {\doibase
  10.1038/s41586-019-1709-y} {\bibfield  {journal} {\bibinfo  {journal}
  {Nature}\ }\textbf {\bibinfo {volume} {575}},\ \bibinfo {pages} {622--627}
  (\bibinfo {year} {2019})}\BibitemShut {NoStop}%
\bibitem [{\citenamefont {Senellart}, \citenamefont {Solomon},\ and\
  \citenamefont {White}(2017)}]{Senellart2017}%
  \BibitemOpen
  \bibfield  {author} {\bibinfo {author} {\bibfnamefont {P.}~\bibnamefont
  {Senellart}}, \bibinfo {author} {\bibfnamefont {G.}~\bibnamefont {Solomon}},
  \ and\ \bibinfo {author} {\bibfnamefont {A.}~\bibnamefont {White}},\
  }\bibfield  {title} {\enquote {\bibinfo {title} {High-performance
  semiconductor quantum-dot single-photon sources},}\ }\href {\doibase
  10.1038/NNANO.2017.218} {\bibfield  {journal} {\bibinfo  {journal} {Nat.
  Nanotechnol.}\ }\textbf {\bibinfo {volume} {12}},\ \bibinfo {pages}
  {1026--1039} (\bibinfo {year} {2017})}\BibitemShut {NoStop}%
\bibitem [{\citenamefont {Demanet}\ and\ \citenamefont
  {Marais}(1985)}]{Demanet1985}%
  \BibitemOpen
  \bibfield  {author} {\bibinfo {author} {\bibfnamefont {C.~M.}\ \bibnamefont
  {Demanet}}\ and\ \bibinfo {author} {\bibfnamefont {M.~A.}\ \bibnamefont
  {Marais}},\ }\bibfield  {title} {\enquote {\bibinfo {title} {A multilayer
  model for {GaAs} oxides formed at room temperature in air as deduced from an
  {XPS} analysis},}\ }\href {\doibase https://doi.org/10.1002/sia.740070104}
  {\bibfield  {journal} {\bibinfo  {journal} {Surf. Interface Anal.}\ }\textbf
  {\bibinfo {volume} {7}},\ \bibinfo {pages} {13--16} (\bibinfo {year}
  {1985})}\BibitemShut {NoStop}%
\bibitem [{\citenamefont {Tomm}\ \emph {et~al.}(2021)\citenamefont {Tomm},
  \citenamefont {Javadi}, \citenamefont {Antoniadis}, \citenamefont {Najer},
  \citenamefont {L\"{o}bl}, \citenamefont {Korsch}, \citenamefont {Schott},
  \citenamefont {Valentin}, \citenamefont {Wieck}, \citenamefont {Ludwig},\
  and\ \citenamefont {Warburton}}]{Tomm2020}%
  \BibitemOpen
  \bibfield  {author} {\bibinfo {author} {\bibfnamefont {N.}~\bibnamefont
  {Tomm}}, \bibinfo {author} {\bibfnamefont {A.}~\bibnamefont {Javadi}},
  \bibinfo {author} {\bibfnamefont {N.~O.}\ \bibnamefont {Antoniadis}},
  \bibinfo {author} {\bibfnamefont {D.}~\bibnamefont {Najer}}, \bibinfo
  {author} {\bibfnamefont {M.~C.}\ \bibnamefont {L\"{o}bl}}, \bibinfo {author}
  {\bibfnamefont {A.~R.}\ \bibnamefont {Korsch}}, \bibinfo {author}
  {\bibfnamefont {R.}~\bibnamefont {Schott}}, \bibinfo {author} {\bibfnamefont
  {S.~R.}\ \bibnamefont {Valentin}}, \bibinfo {author} {\bibfnamefont {A.~D.}\
  \bibnamefont {Wieck}}, \bibinfo {author} {\bibfnamefont {A.}~\bibnamefont
  {Ludwig}}, \ and\ \bibinfo {author} {\bibfnamefont {R.~J.}\ \bibnamefont
  {Warburton}},\ }\bibfield  {title} {\enquote {\bibinfo {title} {A bright and
  fast source of coherent single photons},}\ }\href@noop {} {\bibfield
  {journal} {\bibinfo  {journal} {Nat. Nanotechnol.}\ } (\bibinfo {year}
  {2021})}\BibitemShut {NoStop}%
\bibitem [{\citenamefont {Casey}, \citenamefont {Sell},\ and\ \citenamefont
  {Wecht}(1975)}]{Casey1975}%
  \BibitemOpen
  \bibfield  {author} {\bibinfo {author} {\bibfnamefont {H.~C.}\ \bibnamefont
  {Casey}}, \bibinfo {author} {\bibfnamefont {D.~D.}\ \bibnamefont {Sell}}, \
  and\ \bibinfo {author} {\bibfnamefont {K.~W.}\ \bibnamefont {Wecht}},\
  }\bibfield  {title} {\enquote {\bibinfo {title} {Concentration dependence of
  the absorption coefficient for n- and p-type {G}a{A}s between 1.3 and 1.6
  e{V}},}\ }\href@noop {} {\bibfield  {journal} {\bibinfo  {journal} {J. Appl.
  Phys.}\ }\textbf {\bibinfo {volume} {46}},\ \bibinfo {pages} {250--257}
  (\bibinfo {year} {1975})}\BibitemShut {NoStop}%
\bibitem [{\citenamefont {Franz}(1958)}]{Franz1958}%
  \BibitemOpen
  \bibfield  {author} {\bibinfo {author} {\bibfnamefont {W.}~\bibnamefont
  {Franz}},\ }\bibfield  {title} {\enquote {\bibinfo {title} {Einflu{\ss} eines
  elektrischen {Feldes} auf eine optische {Absorptionskante}},}\ }\href
  {https://www.degruyter.com/view/j/zna.1958.13.issue-6/zna-1958-0609/zna-1958-0609.xml}
  {\bibfield  {journal} {\bibinfo  {journal} {Zeitschrift für Naturforschung
  A}\ }\textbf {\bibinfo {volume} {13}},\ \bibinfo {pages} {484} (\bibinfo
  {year} {1958})}\BibitemShut {NoStop}%
\bibitem [{\citenamefont {Keldysh}(1957)}]{Keldysh1957}%
  \BibitemOpen
  \bibfield  {author} {\bibinfo {author} {\bibfnamefont {V.~L.}\ \bibnamefont
  {Keldysh}},\ }\bibfield  {title} {\enquote {\bibinfo {title} {Behaviour of
  non-metallic crystals in strong electric fields},}\ }\href
  {http://www.jetp.ac.ru/cgi-bin/e/index/e/6/4/p763?a=list} {\bibfield
  {journal} {\bibinfo  {journal} {Journal of Experimental and Theoretical
  Physics (USSR)}\ }\textbf {\bibinfo {volume} {33}},\ \bibinfo {pages} {994}
  (\bibinfo {year} {1957})}\BibitemShut {NoStop}%
\bibitem [{\citenamefont {Aspnes}(1966)}]{Aspnes1966}%
  \BibitemOpen
  \bibfield  {author} {\bibinfo {author} {\bibfnamefont {D.~E.}\ \bibnamefont
  {Aspnes}},\ }\bibfield  {title} {\enquote {\bibinfo {title} {Electric-field
  effects on optical absorption near thresholds in solids},}\ }\href {\doibase
  10.1103/PhysRev.147.554} {\bibfield  {journal} {\bibinfo  {journal} {Phys.
  Rev.}\ }\textbf {\bibinfo {volume} {147}},\ \bibinfo {pages} {554--566}
  (\bibinfo {year} {1966})}\BibitemShut {NoStop}%
\bibitem [{\citenamefont {Hader}, \citenamefont {Linder},\ and\ \citenamefont
  {D\"ohler}(1997)}]{Hader1997}%
  \BibitemOpen
  \bibfield  {author} {\bibinfo {author} {\bibfnamefont {J.}~\bibnamefont
  {Hader}}, \bibinfo {author} {\bibfnamefont {N.}~\bibnamefont {Linder}}, \
  and\ \bibinfo {author} {\bibfnamefont {G.~H.}\ \bibnamefont {D\"ohler}},\
  }\bibfield  {title} {\enquote {\bibinfo {title} {k\ensuremath{\cdot}p theory
  of the {F}ranz-{K}eldysh effect},}\ }\href {\doibase
  10.1103/PhysRevB.55.6960} {\bibfield  {journal} {\bibinfo  {journal} {Phys.
  Rev. B}\ }\textbf {\bibinfo {volume} {55}},\ \bibinfo {pages} {6960--6974}
  (\bibinfo {year} {1997})}\BibitemShut {NoStop}%
\bibitem [{\citenamefont {{Knupfer}}\ \emph {et~al.}(1993)\citenamefont
  {{Knupfer}}, \citenamefont {{Kiesel}}, \citenamefont {{Kneissl}},
  \citenamefont {{Dankowski}}, \citenamefont {{Linder}}, \citenamefont
  {{Weimann}},\ and\ \citenamefont {{Dohler}}}]{Knupfer1993}%
  \BibitemOpen
  \bibfield  {author} {\bibinfo {author} {\bibfnamefont {B.}~\bibnamefont
  {{Knupfer}}}, \bibinfo {author} {\bibfnamefont {P.}~\bibnamefont {{Kiesel}}},
  \bibinfo {author} {\bibfnamefont {M.}~\bibnamefont {{Kneissl}}}, \bibinfo
  {author} {\bibfnamefont {S.}~\bibnamefont {{Dankowski}}}, \bibinfo {author}
  {\bibfnamefont {N.}~\bibnamefont {{Linder}}}, \bibinfo {author}
  {\bibfnamefont {G.}~\bibnamefont {{Weimann}}}, \ and\ \bibinfo {author}
  {\bibfnamefont {G.~H.}\ \bibnamefont {{Dohler}}},\ }\bibfield  {title}
  {\enquote {\bibinfo {title} {Polarization-insensitive high-contrast
  {G}a{A}s/{A}l{G}a{A}s waveguide modulator based on the {F}ranz-{K}eldysh
  effect},}\ }\href {\doibase 10.1109/68.262549} {\bibfield  {journal}
  {\bibinfo  {journal} {IEEE Photonics Technology Letters}\ }\textbf {\bibinfo
  {volume} {5}},\ \bibinfo {pages} {1386--1388} (\bibinfo {year}
  {1993})}\BibitemShut {NoStop}%
\bibitem [{\citenamefont {Greuter}\ \emph {et~al.}(2014)\citenamefont
  {Greuter}, \citenamefont {Starosielec}, \citenamefont {Najer}, \citenamefont
  {Ludwig}, \citenamefont {Duempelmann}, \citenamefont {Rohner},\ and\
  \citenamefont {Warburton}}]{Greuter2014}%
  \BibitemOpen
  \bibfield  {author} {\bibinfo {author} {\bibfnamefont {L.}~\bibnamefont
  {Greuter}}, \bibinfo {author} {\bibfnamefont {S.}~\bibnamefont
  {Starosielec}}, \bibinfo {author} {\bibfnamefont {D.}~\bibnamefont {Najer}},
  \bibinfo {author} {\bibfnamefont {A.}~\bibnamefont {Ludwig}}, \bibinfo
  {author} {\bibfnamefont {L.}~\bibnamefont {Duempelmann}}, \bibinfo {author}
  {\bibfnamefont {D.}~\bibnamefont {Rohner}}, \ and\ \bibinfo {author}
  {\bibfnamefont {R.~J.}\ \bibnamefont {Warburton}},\ }\bibfield  {title}
  {\enquote {\bibinfo {title} {A small mode volume tunable microcavity:
  Development and characterization},}\ }\href@noop {} {\bibfield  {journal}
  {\bibinfo  {journal} {Appl. Phys. Lett.}\ }\textbf {\bibinfo {volume}
  {105}},\ \bibinfo {pages} {121105} (\bibinfo {year} {2014})}\BibitemShut
  {NoStop}%
\bibitem [{\citenamefont {Hunger}\ \emph {et~al.}(2012)\citenamefont {Hunger},
  \citenamefont {Deutsch}, \citenamefont {Barbour}, \citenamefont {Warburton},\
  and\ \citenamefont {Reichel}}]{Hunger2012}%
  \BibitemOpen
  \bibfield  {author} {\bibinfo {author} {\bibfnamefont {D.}~\bibnamefont
  {Hunger}}, \bibinfo {author} {\bibfnamefont {C.}~\bibnamefont {Deutsch}},
  \bibinfo {author} {\bibfnamefont {R.~J.}\ \bibnamefont {Barbour}}, \bibinfo
  {author} {\bibfnamefont {R.~J.}\ \bibnamefont {Warburton}}, \ and\ \bibinfo
  {author} {\bibfnamefont {J.}~\bibnamefont {Reichel}},\ }\bibfield  {title}
  {\enquote {\bibinfo {title} {Laser micro-fabrication of concave,
  low-roughness features in silica},}\ }\href@noop {} {\bibfield  {journal}
  {\bibinfo  {journal} {AIP Adv.}\ }\textbf {\bibinfo {volume} {2}},\ \bibinfo
  {pages} {012119} (\bibinfo {year} {2012})}\BibitemShut {NoStop}%
\bibitem [{Note1()}]{Note1}%
  \BibitemOpen
  \bibinfo {note} {Note that in this work, we define the SB center as the mean
  value of the two wavelengths at the local minima (with $R<90\%$) of the
  calculated reflectance spectrum that are closest to the maximum mirror
  reflectance (Fig.~\ref {design2}(b)).}\BibitemShut {Stop}%
\bibitem [{\citenamefont {Warburton}\ \emph {et~al.}(2000)\citenamefont
  {Warburton}, \citenamefont {Schaflein}, \citenamefont {Haft}, \citenamefont
  {Bickel}, \citenamefont {Lorke}, \citenamefont {Karrai}, \citenamefont
  {Garcia}, \citenamefont {Schoenfeld},\ and\ \citenamefont
  {Petroff}}]{Warburton2000}%
  \BibitemOpen
  \bibfield  {author} {\bibinfo {author} {\bibfnamefont {R.~J.}\ \bibnamefont
  {Warburton}}, \bibinfo {author} {\bibfnamefont {C.}~\bibnamefont
  {Schaflein}}, \bibinfo {author} {\bibfnamefont {D.}~\bibnamefont {Haft}},
  \bibinfo {author} {\bibfnamefont {F.}~\bibnamefont {Bickel}}, \bibinfo
  {author} {\bibfnamefont {A.}~\bibnamefont {Lorke}}, \bibinfo {author}
  {\bibfnamefont {K.}~\bibnamefont {Karrai}}, \bibinfo {author} {\bibfnamefont
  {J.~M.}\ \bibnamefont {Garcia}}, \bibinfo {author} {\bibfnamefont
  {W.}~\bibnamefont {Schoenfeld}}, \ and\ \bibinfo {author} {\bibfnamefont
  {P.~M.}\ \bibnamefont {Petroff}},\ }\bibfield  {title} {\enquote {\bibinfo
  {title} {Optical emission from a charge-tunable quantum ring},}\ }\href
  {\doibase 10.1038/35016030} {\bibfield  {journal} {\bibinfo  {journal}
  {Nature}\ }\textbf {\bibinfo {volume} {405}},\ \bibinfo {pages} {926--929}
  (\bibinfo {year} {2000})}\BibitemShut {NoStop}%
\bibitem [{\citenamefont {Liu}\ \emph {et~al.}(2018)\citenamefont {Liu},
  \citenamefont {Konthasinghe}, \citenamefont {Davan\ifmmode~\mbox{\c{c}}\else
  \c{c}\fi{}o}, \citenamefont {Lawall}, \citenamefont {Anant}, \citenamefont
  {Verma}, \citenamefont {Mirin}, \citenamefont {Nam}, \citenamefont {Song},
  \citenamefont {Ma}, \citenamefont {Chen}, \citenamefont {Ni}, \citenamefont
  {Niu},\ and\ \citenamefont {Srinivasan}}]{Liu2018}%
  \BibitemOpen
  \bibfield  {author} {\bibinfo {author} {\bibfnamefont {J.}~\bibnamefont
  {Liu}}, \bibinfo {author} {\bibfnamefont {K.}~\bibnamefont {Konthasinghe}},
  \bibinfo {author} {\bibfnamefont {M.}~\bibnamefont
  {Davan\ifmmode~\mbox{\c{c}}\else \c{c}\fi{}o}}, \bibinfo {author}
  {\bibfnamefont {J.}~\bibnamefont {Lawall}}, \bibinfo {author} {\bibfnamefont
  {V.}~\bibnamefont {Anant}}, \bibinfo {author} {\bibfnamefont
  {V.}~\bibnamefont {Verma}}, \bibinfo {author} {\bibfnamefont
  {R.}~\bibnamefont {Mirin}}, \bibinfo {author} {\bibfnamefont {S.~W.}\
  \bibnamefont {Nam}}, \bibinfo {author} {\bibfnamefont {J.~D.}\ \bibnamefont
  {Song}}, \bibinfo {author} {\bibfnamefont {B.}~\bibnamefont {Ma}}, \bibinfo
  {author} {\bibfnamefont {Z.~S.}\ \bibnamefont {Chen}}, \bibinfo {author}
  {\bibfnamefont {H.~Q.}\ \bibnamefont {Ni}}, \bibinfo {author} {\bibfnamefont
  {Z.~C.}\ \bibnamefont {Niu}}, \ and\ \bibinfo {author} {\bibfnamefont
  {K.}~\bibnamefont {Srinivasan}},\ }\bibfield  {title} {\enquote {\bibinfo
  {title} {Single self-assembled $\mathrm{InAs}/\mathrm{GaAs}$ quantum dots in
  photonic nanostructures: The role of nanofabrication},}\ }\href {\doibase
  10.1103/PhysRevApplied.9.064019} {\bibfield  {journal} {\bibinfo  {journal}
  {Phys. Rev. Appl.}\ }\textbf {\bibinfo {volume} {9}},\ \bibinfo {pages}
  {064019} (\bibinfo {year} {2018})}\BibitemShut {NoStop}%
\bibitem [{\citenamefont {Xuan}, \citenamefont {Lin},\ and\ \citenamefont
  {Ye}(2007)}]{Xuan2007}%
  \BibitemOpen
  \bibfield  {author} {\bibinfo {author} {\bibfnamefont {Y.}~\bibnamefont
  {Xuan}}, \bibinfo {author} {\bibfnamefont {H.}~\bibnamefont {Lin}}, \ and\
  \bibinfo {author} {\bibfnamefont {P.~D.}\ \bibnamefont {Ye}},\ }\bibfield
  {title} {\enquote {\bibinfo {title} {Simplified surface preparation for
  {GaAs} passivation using atomic layer-deposited high-$\kappa$ dielectrics},}\
  }\href {\doibase 10.1109/TED.2007.900678} {\bibfield  {journal} {\bibinfo
  {journal} {IEEE Transactions on Electron Devices}\ }\textbf {\bibinfo
  {volume} {54}},\ \bibinfo {pages} {1811--1817} (\bibinfo {year}
  {2007})}\BibitemShut {NoStop}%
\bibitem [{\citenamefont {Rebaud}\ \emph {et~al.}(2015)\citenamefont {Rebaud},
  \citenamefont {Roure}, \citenamefont {Loup}, \citenamefont {Rodriguez},
  \citenamefont {Martinez},\ and\ \citenamefont {Besson}}]{Rebaud2015}%
  \BibitemOpen
  \bibfield  {author} {\bibinfo {author} {\bibfnamefont {M.}~\bibnamefont
  {Rebaud}}, \bibinfo {author} {\bibfnamefont {M.-C.}\ \bibnamefont {Roure}},
  \bibinfo {author} {\bibfnamefont {V.}~\bibnamefont {Loup}}, \bibinfo {author}
  {\bibfnamefont {P.}~\bibnamefont {Rodriguez}}, \bibinfo {author}
  {\bibfnamefont {E.}~\bibnamefont {Martinez}}, \ and\ \bibinfo {author}
  {\bibfnamefont {P.}~\bibnamefont {Besson}},\ }\bibfield  {title} {\enquote
  {\bibinfo {title} {Chemical treatments for native oxides removal of {GaAs}
  wafers},}\ }\href {\doibase 10.1149/06908.0243ecst} {\bibfield  {journal}
  {\bibinfo  {journal} {{ECS} Trans.}\ }\textbf {\bibinfo {volume} {69}},\
  \bibinfo {pages} {243--250} (\bibinfo {year} {2015})}\BibitemShut {NoStop}%
\bibitem [{\citenamefont {Yablonovitch}\ \emph {et~al.}(1987)\citenamefont
  {Yablonovitch}, \citenamefont {Sandroff}, \citenamefont {Bhat},\ and\
  \citenamefont {Gmitter}}]{Yablonovitch1987}%
  \BibitemOpen
  \bibfield  {author} {\bibinfo {author} {\bibfnamefont {E.}~\bibnamefont
  {Yablonovitch}}, \bibinfo {author} {\bibfnamefont {C.~J.}\ \bibnamefont
  {Sandroff}}, \bibinfo {author} {\bibfnamefont {R.}~\bibnamefont {Bhat}}, \
  and\ \bibinfo {author} {\bibfnamefont {T.}~\bibnamefont {Gmitter}},\
  }\bibfield  {title} {\enquote {\bibinfo {title} {Nearly ideal electronic
  properties of sulfide coated {GaAs} surfaces},}\ }\href {\doibase
  10.1063/1.98415} {\bibfield  {journal} {\bibinfo  {journal} {Appl. Phys.
  Lett.}\ }\textbf {\bibinfo {volume} {51}},\ \bibinfo {pages} {439--441}
  (\bibinfo {year} {1987})}\BibitemShut {NoStop}%
\bibitem [{\citenamefont {Ohno}(1991)}]{Ohno1991}%
  \BibitemOpen
  \bibfield  {author} {\bibinfo {author} {\bibfnamefont {T.}~\bibnamefont
  {Ohno}},\ }\bibfield  {title} {\enquote {\bibinfo {title} {Sulfur passivation
  of {GaAs} surfaces},}\ }\href {\doibase 10.1103/PhysRevB.44.6306} {\bibfield
  {journal} {\bibinfo  {journal} {Phys. Rev. B}\ }\textbf {\bibinfo {volume}
  {44}},\ \bibinfo {pages} {6306--6311} (\bibinfo {year} {1991})}\BibitemShut
  {NoStop}%
\bibitem [{\citenamefont {Kuhlmann}\ \emph {et~al.}(2013)\citenamefont
  {Kuhlmann}, \citenamefont {Houel}, \citenamefont {Ludwig}, \citenamefont
  {Greuter}, \citenamefont {Reuter}, \citenamefont {Wieck}, \citenamefont
  {Poggio},\ and\ \citenamefont {Warburton}}]{Kuhlmann2013}%
  \BibitemOpen
  \bibfield  {author} {\bibinfo {author} {\bibfnamefont {A.~V.}\ \bibnamefont
  {Kuhlmann}}, \bibinfo {author} {\bibfnamefont {J.}~\bibnamefont {Houel}},
  \bibinfo {author} {\bibfnamefont {A.}~\bibnamefont {Ludwig}}, \bibinfo
  {author} {\bibfnamefont {L.}~\bibnamefont {Greuter}}, \bibinfo {author}
  {\bibfnamefont {D.}~\bibnamefont {Reuter}}, \bibinfo {author} {\bibfnamefont
  {A.~D.}\ \bibnamefont {Wieck}}, \bibinfo {author} {\bibfnamefont
  {M.}~\bibnamefont {Poggio}}, \ and\ \bibinfo {author} {\bibfnamefont {R.~J.}\
  \bibnamefont {Warburton}},\ }\bibfield  {title} {\enquote {\bibinfo {title}
  {Charge noise and spin noise in a semiconductor quantum device},}\ }\href
  {\doibase 10.1038/NPHYS2688} {\bibfield  {journal} {\bibinfo  {journal} {Nat.
  Phys.}\ }\textbf {\bibinfo {volume} {9}},\ \bibinfo {pages} {570--575}
  (\bibinfo {year} {2013})}\BibitemShut {NoStop}%
\bibitem [{\citenamefont {Robertson}, \citenamefont {Guo},\ and\ \citenamefont
  {Lin}(2015)}]{Robertson2015}%
  \BibitemOpen
  \bibfield  {author} {\bibinfo {author} {\bibfnamefont {J.}~\bibnamefont
  {Robertson}}, \bibinfo {author} {\bibfnamefont {Y.}~\bibnamefont {Guo}}, \
  and\ \bibinfo {author} {\bibfnamefont {L.}~\bibnamefont {Lin}},\ }\bibfield
  {title} {\enquote {\bibinfo {title} {Defect state passivation at {III-V}
  oxide interfaces for complementary metal–oxide–semiconductor devices},}\
  }\href {\doibase 10.1063/1.4913832} {\bibfield  {journal} {\bibinfo
  {journal} {J. Appl. Phys.}\ }\textbf {\bibinfo {volume} {117}},\ \bibinfo
  {pages} {112806} (\bibinfo {year} {2015})}\BibitemShut {NoStop}%
\bibitem [{Note2()}]{Note2}%
  \BibitemOpen
  \bibinfo {note} {Note that the measured $\protect \mathcal {Q}$-factors
  obtained with an electrically-contacted passivated sample and with an
  electrically-uncontacted passivated sample were similar. The latter are not
  shown in Fig.~\ref {cavitycharac}(b).}\BibitemShut {Stop}%
\bibitem [{\citenamefont {Carniglia}\ and\ \citenamefont
  {Jensen}(2002)}]{Carniglia2002}%
  \BibitemOpen
  \bibfield  {author} {\bibinfo {author} {\bibfnamefont {C.~K.}\ \bibnamefont
  {Carniglia}}\ and\ \bibinfo {author} {\bibfnamefont {D.~G.}\ \bibnamefont
  {Jensen}},\ }\bibfield  {title} {\enquote {\bibinfo {title} {Single-layer
  model for surface roughness},}\ }\href {\doibase 10.1364/AO.41.003167}
  {\bibfield  {journal} {\bibinfo  {journal} {Appl. Opt.}\ }\textbf {\bibinfo
  {volume} {41}},\ \bibinfo {pages} {3167--3171} (\bibinfo {year}
  {2002})}\BibitemShut {NoStop}%
\bibitem [{Note3()}]{Note3}%
  \BibitemOpen
  \bibinfo {note} {We note that the semiconductor heterostructure without
  doping was not grown with the same MBE as the semiconductor heterostructure
  with doping.}\BibitemShut {Stop}%
\bibitem [{\citenamefont {Bennett}(1992)}]{Bennett1992}%
  \BibitemOpen
  \bibfield  {author} {\bibinfo {author} {\bibfnamefont {J.~M.}\ \bibnamefont
  {Bennett}},\ }\bibfield  {title} {\enquote {\bibinfo {title} {Recent
  developments in surface roughness characterization},}\ }\href {\doibase
  10.1088/0957-0233/3/12/001} {\bibfield  {journal} {\bibinfo  {journal} {Meas.
  Sci. Technol.}\ }\textbf {\bibinfo {volume} {3}},\ \bibinfo {pages}
  {1119--1127} (\bibinfo {year} {1992})}\BibitemShut {NoStop}%
\bibitem [{\citenamefont {Davies}(1997)}]{Davies1998}%
  \BibitemOpen
  \bibfield  {author} {\bibinfo {author} {\bibfnamefont {J.}~\bibnamefont
  {Davies}},\ }\href
  {https://www.cambridge.org/core/books/physics-of-lowdimensional-semiconductors/D1B7DE285E09FCA518C4C6C1C385E466}
  {\emph {\bibinfo {title} {The Physics of Low-dimensional Semiconductors: An
  Introduction}}}\ (\bibinfo  {publisher} {Cambridge University Press},\
  \bibinfo {year} {1997})\BibitemShut {NoStop}%
\bibitem [{Note4()}]{Note4}%
  \BibitemOpen
  \bibinfo {note} {The Airy function is defined as ${\protect \rm
  Ai}(z)=\protect \frac {1}{2\pi }\DOTSI \intop \ilimits@ _{-\infty }^{\infty
  }e^{i(zt+t^3/3)}dt$.}\BibitemShut {Stop}%
\bibitem [{\citenamefont {Zamora~Peredo}\ \emph {et~al.}(2016)\citenamefont
  {Zamora~Peredo}, \citenamefont {García-González}, \citenamefont
  {Hernandez~Torres}, \citenamefont {Cortes-Mestizo}, \citenamefont
  {Mendez-Garcia},\ and\ \citenamefont {López-López}}]{Zamora2016}%
  \BibitemOpen
  \bibfield  {author} {\bibinfo {author} {\bibfnamefont {L.}~\bibnamefont
  {Zamora~Peredo}}, \bibinfo {author} {\bibfnamefont {L.}~\bibnamefont
  {García-González}}, \bibinfo {author} {\bibfnamefont {J.}~\bibnamefont
  {Hernandez~Torres}}, \bibinfo {author} {\bibfnamefont {I.}~\bibnamefont
  {Cortes-Mestizo}}, \bibinfo {author} {\bibfnamefont {V.}~\bibnamefont
  {Mendez-Garcia}}, \ and\ \bibinfo {author} {\bibfnamefont {M.}~\bibnamefont
  {López-López}},\ }\bibfield  {title} {\enquote {\bibinfo {title}
  {Photoreflectance and {R}aman study of surface electric states on
  {AlGaAs}/{GaAs} heterostructures},}\ }\href {\doibase 10.1155/2016/4601249}
  {\bibfield  {journal} {\bibinfo  {journal} {J. Spectrosc.}\ }\textbf
  {\bibinfo {volume} {2016}},\ \bibinfo {pages} {1--8} (\bibinfo {year}
  {2016})}\BibitemShut {NoStop}%
\bibitem [{Note5()}]{Note5}%
  \BibitemOpen
  \bibinfo {note} {In order to estimate the capping fields in Ref.~\cite
  {Zamora2016}, we assume mid-gap pinning, dividing half the bandgap
  $E_{\protect \rm g}/2=0.71$\protect \tmspace +\thinmuskip {.1667em}eV of GaAs
  at 300\protect \tmspace +\thinmuskip {.1667em}K by the capping layer
  thicknesses there reported.}\BibitemShut {Stop}%
\bibitem [{Note6()}]{Note6}%
  \BibitemOpen
  \bibinfo {note} {We note that reflectance spectra (Fig.~\ref
  {mirrorcharac}(a)) on different samples with nominally the same dielectric
  coating exhibited up to 6\protect \tmspace +\thinmuskip {.1667em}nm shifts in
  wavelength, most probably due to thickness variations across the
  wafer.}\BibitemShut {Stop}%
\bibitem [{\citenamefont {Beauville}\ and\ \citenamefont {{The {VIRGO}
  Collaboration}}(2004)}]{BeauvilleCQG2004}%
  \BibitemOpen
  \bibfield  {author} {\bibinfo {author} {\bibfnamefont {F.}~\bibnamefont
  {Beauville}}\ and\ \bibinfo {author} {\bibnamefont {{The {VIRGO}
  Collaboration}}},\ }\bibfield  {title} {\enquote {\bibinfo {title} {The
  {VIRGO} large mirrors: a challenge for low loss coatings},}\ }\href {\doibase
  10.1088/0264-9381/21/5/083} {\bibfield  {journal} {\bibinfo  {journal}
  {Class. Quantum Gravity}\ }\textbf {\bibinfo {volume} {21}},\ \bibinfo
  {pages} {S935--S945} (\bibinfo {year} {2004})}\BibitemShut {NoStop}%
\end{thebibliography}%
\end{document}